\documentclass[12pt,preprint]{aastex}

\shorttitle{Triggered Star Formation}

\shortauthors{H. T. Lee and W. P. Chen}

\begin{document}

\title{TRIGGERED STAR FORMATION BY MASSIVE STARS}

\author{Hsu-Tai Lee\altaffilmark{1} \email{htlee@asiaa.sinica.edu.tw}}

\author{W. P. Chen\altaffilmark{1,2} \email{wchen@astro.ncu.edu.tw}}

\altaffiltext{1}{Institute of Astronomy, National Central University, 300 Jungda Road, Jungli 
32054, Taiwan}

\altaffiltext{2}{Department of Physics, National Central University, 300 Jungda Road, Jungli  
32054, Taiwan}

\begin{abstract}
We present our diagnosis of the role that massive stars play in the formation of low- and 
intermediate-mass stars in OB associations (the $\lambda$ Ori region, the Ori\,OB1, and Lac\,OB1 
associations).  We find that the classical T Tauri stars and Herbig Ae/Be stars tend to 
line up between luminous O stars and bright-rimmed or comet-shaped clouds; the closer to a cloud 
the progressively younger they are.  Our positional and chronological study lends support to the 
validity of the radiation-driven implosion mechanism, where the Lyman continuum photons from a 
luminous O star create expanding ionization fronts to evaporate and 
compress nearby clouds into bright-rimmed or comet-shaped clouds.  Implosive pressure then 
causes dense clumps to collapse, prompting the formation of low-mass stars on the cloud surface 
(i.e., the bright rim) and intermediate-mass stars somewhat deeper into the cloud.  These stars 
are a signpost of current star formation; no young stars are seen leading the ionization 
fronts further into the cloud.  Young stars in bright-rimmed or comet-shaped clouds 
are likely to have been formed by triggering, which would result in an age spread of several 
Myrs between the member stars or star groups formed in the sequence.
\end{abstract}


\section{INTRODUCTION}

Most O and B stars are congregated into OB associations \citep{bla64} in which young low- 
(classical T Tauri stars, or CTTSs) and intermediate-mass (Herbig Ae/Be stars, or HAeBe) stellar 
groups are also found \citep[see the review by][]{bri06}.  What is the relationship between the 
formation of massive stars and that of low-mass stars?  Does star formation in an OB association 
proceed in a bimodal manner for massive and for low-mass stellar groups?  If so, which group 
would form first?  It is noted that massive stars have a profound influence on the surrounding 
molecular clouds.  On the one hand, the radiation and energetic wind from a massive star could 
cause the evaporation of nearby clouds, hence terminating the star formation processes.  On the 
other hand, the massive star could provide ``just the touch'' needed to prompt the collapse of a 
molecular cloud which otherwise may not contract and fragment spontaneously.  Do massive stars 
play primarily a destructive or promotional role in star formation in a molecular 
cloud?  \citet{her62} suggests that low- and intermediate-mass stars form first in an OB 
association, but soon after massive O stars appear, the cloud is disrupted, which hinders further 
star formation.  Alternatively, \citet{elm77} and \citet{lad87} propose that low-mass stars form 
first out of cloud fragments and are distributed throughout the entire molecular cloud.  Once the 
O stars form, their expanding ionization fronts (I-fronts) then play a constructive role in 
inciting a sequence of star formation in neighboring molecular clouds.

The triggering of star formation by massive stars appears to take place on different length scales 
\citep{elm98}.  The Sco\,OB2 association might be one example of triggered star formation 
\citep{deg89}.  In this case the Upper Centaurus Lupus subgroup was formed first in the middle of 
the molecular cloud complex, which then prompted star formation on both sides, eventually becoming 
the Upper Scorpius and Lower Centaurus Crux associations.  \citet{pre99,pre06} propose a similar 
mechanism, but with a series of supernova explosions as the triggering sources.  The star formation 
activities can be sustained as long as stars massive enough are produced in the sequence 
and there is enough surrounding material.  This sequential formation process leads naturally to an 
age spread among member stars or subgroups \citep{bla64}, and the stellar aggregates thus formed 
(out of separate clouds) tend to be sparsely distributed and gravitationally unbound because of 
the expanding I-fronts or an initially unbound giant molecular cloud  \citep[see][]{cla05}.

On a smaller scale, the signature of ongoing star formation, such as young stellar jets, 
evaporated gaseous globules (EGGs) and water masers have been found in the periphery of  
\ion{H}{2} regions \citep{hes04,hes05}.   There are two kinds of triggering mechanisms 
\citep[e.g.,][]{kar03}, ``collect-and-collapse" and radiation-driven implosion (RDI).  In the 
collect-and-collapse scenario first proposed by \citet{elm77} and recently demonstrated 
observationally by \citet{deh05}, \citet{zav06}, and \citet{san06}, the expanding 
I-fronts from an \ion{H}{2} region piles up a shell of dense gas and dust, in which clumps 
fragment and collapse to form the next generation of stars.  In the RDI scenario 
\citep{ber89,ber90,hes05,lar83,kes03}, the formation sequence begins with photoionization of a 
nearby molecular cloud by a massive star.  The shock fronts embracing the surface of the cloud 
compress the cloud until reaching the critical density for gravitational collapse resulting in the 
formation of new stars.  The latest star formation, as traced by protostellar cores \citep{lef00} 
or water masers \citep{hea04}, takes place at the compressed layer of a cloud.  \citet{hes05} 
propose a scenario in which an EGG appears when a dense clump is impinged upon by the I-fronts.  The 
photoevaporation then erodes the circumstellar disk into a protoplanetary disk, or a ``proplyd'' 
\citep{ode93}.  Subsequently formed massive stars can carve out their own cavities to continue the 
triggering process \citep{san06}.  The exposure of the protoplanetary disk in such 
environments would, in addition to being truncated in size, contain short-lived radio nuclides 
from the ejecta from one or more nearby supernovae, such as has been observed in meteorites in the 
Solar System \citep{hes04}.  Our results substantiate the above sequential star formation scenario by 
providing clear chronological and positional evidence that massive stars prompt the birth of lower-mass stars out 
of molecular clouds.

In the Orion star-forming region for example, there is concrete evidence of triggered star 
formation as manifested by the bright-rimmed clouds (BRCs) in the vicinity of O stars 
\citep[hereafter Paper~I]{lee05}.  These BRCs are considered the remnant of molecular clouds that 
have been photoionized by a nearby massive star \citep{sug91,sug94}.  According to Paper~I, only 
BRCs that are associated with strong $IRAS$ 100~$\micron$ emission (tracer of high density) and 
H$\alpha$ emission (tracer of the ionization front) show signs of ongoing star formation.  
Furthermore, CTTSs are more likely seen between the O stars and the BRCs, with those closer 
to the BRCs being progressively younger, and there are no CTTSs far ahead of the I-fronts.  


In this paper we extend the study to the Lac\,OB1 association as well as include intermediate-mass 
young stars in our sample.  We describe in \S 2 the archival data and our spectroscopic and 
imaging observations.  In addition to the Lac\,OB1 sources, some of the stars in Ori\,OB1 
considered in Paper~I to be young star candidates have been spectroscopically confirmed.  These 
results are also presented here.  Finally, we discuss star-formation activities and histories in 
$\lambda$ Ori, Ori\,OB1, and Lac\,OB1 in \S 3, and consider star formation in general in OB 
associations in \S 4.  The conclusions are summarized in \S 5.

\section{DATA AND OBSERVATIONS}

\subsection{ARCHIVE DATA}

CTTSs are young stellar objects characterized by their infrared excess.  Usually CTTSs are more 
likely to be spatially closer to a star-forming region than are the weak-line T Tauri stars (WTTSs).   
The latter are also PMS stars, but are more evolved than CTTSs in terms of clearing of their inner 
circumstellar disks.  Thus the CTTSs trace more recent star formation.  In Paper~I, we proposed 
an empirical set of criteria to select CTTS candidates from the 2MASS Point Source Catalog \citep{cut03}. 
In this paper, we apply the same selection procedure (e.g., 2MASS 
colors, good photometric qualities, and exclusion of extended sources) but include young 
intermediate-mass stars, the HAeBe stars into our sample.  Different young stellar populations, 
WTTSs, CTTSs, and HAeBe stars occupy distinctly different regions in the 2MASS color-color diagram 
(see paper~I).  The HAeBe stars in general exhibit larger infrared excess than CTTSs do.  
Therefore we select as HAeBe star candidates 2MASS point sources with colors redder than the line 
defined by $(m_{J}-m_{H})-1.7(m_{H}-m_{K})+0.450=0$; CTTS candidates are selected by the same 
method described in paper~I, namely between the two parallel lines, 
$(m_{J}- m_{H})-1.7(m_{H}-m_{K})+0.0976=0$ and $(m_{J}-m_{H})-1.7(m_{H}-m_{K})+0.450=0$, and above 
the dereddened CTTS locus \citep{mey97}, $(m_{J}-m_{H})-0.493(m_{H}-m_{K})-0.439=0$.

Table~1 shows the fields in the $\lambda$ Ori region, Ori\,OB1, and Lac\,OB1 studied in this 
paper, which include 7 BRCs, one comet-shaped cloud and two control regions.  In addition to the 
2MASS point-source database from which we select our CTTS and HAeBe candidates, we also make use 
of the H$\alpha$ emission survey data \citep{fin03,gau01,den98,haf03}, $E(B-V)$ reddening 
\citep{sch98}, $IRAS$ 100~$\micron$, and CO \citep{dam01} emission to trace, respectively, the 
distribution of the ionization fronts, cloud extinction, IR radiation, and molecular clouds with 
respect to the spatial distribution of our young star sample.

\subsection{SPECTROSCOPIC OBSERVATIONS}

The spectra of bright CTTS and HAeBe candidates were taken at the Beijing Astronomical Observatory 
(BAO) and at the Kitt Peak National Observatory (KPNO).  At the BAO, low-dispersion spectra with a 
dispersion of 200~\AA\,mm$^{-1}$, corresponding to 4.8~\AA\,pixel$^{-1}$, were taken with the 
2.16~m optical telescope from 2003 October~31 to November~3, and on 2004 September 5--6.  An 
OMR (Optomechanics Research, Inc.) spectrograph was used with a Tektronix 1024$\times$1024 CCD 
detector covering 4000--9000~\AA.  These spectra were used to confirm the young stellar 
nature (e.g., the H$\alpha$ and other characteristic emission lines) of the PMS star candidates 
selected on the basis of the 2MASS colors. 

Medium-dispersion spectra for a selected set of sample stars were taken with the KPNO 2.1~m 
telescope on 2004 January 2--5.  The GoldCamera spectrometer, with a Ford 3~K$\times$1~K CCD 
with 15~$\micron$ pixels, was used with the grating \#26new, giving a dispersion of 
1.24~\AA\,pixel$^{-1}$.  These medium-dispersion spectra allowed us to identify the lithium 
absorption at 6708~\AA, the spectral signature of a low-mass PMS star.

All the spectroscopic data were processed with the standard NOAO/IRAF packages.  After correction 
for bias and flat-fields, the IRAF package KPNOSLIT was used to extract and to calibrate the 
wavelength and flux of each spectrum.  To check the legitimacy of our selection criteria,  we 
also observed two control fields, in addition to the star-forming clouds.  All the 
fields included in this study are summarized in Table~1.  

\subsection{IMAGING OBSERVATION}

The BRCs were imaged on 2004 November 3--8 using the 1~m telescope at the Lulin Observatory in 
Taiwan (Table~2).  A PI\,1300B (Roper Scientific) CCD camera was used, which has $1340 \times 
1300$ pixels, each 20~$\micron$ square, yielding a $\sim 11\arcmin$ field of view.  H$\alpha$ 
($\lambda_c =6563$~\AA, $\Delta\lambda$(FWHM)=30~\AA) images were taken for all BRCs.  In 
addition, LBN\,437 was observed with an [\ion{S}{2}] ($\lambda_c =6724$~\AA, 
$\Delta\lambda$(FWHM)=80~\AA) filter.  For every target field tens of images were taken, each with 
an exposure time of 120 to 300~s.  The images were processed for bias, dark and flat-fielding 
corrections with the standard procedures.

\subsection{OBSERVATIONAL RESULTS}

The main purpose of the spectral observations was to identify PMS star candidates and to validate the 
selection criteria for HAeBe stars.  The imaging observations can help us to trace the distribution of the 
I-fronts in the BRCs.  By combining the spectral and imaging observations, we can study the spatial 
distribution of PMS stars relative to I-fronts in BRCs.

\begin{figure}
\epsscale{1.0}
\includegraphics[angle=90,scale=0.5]{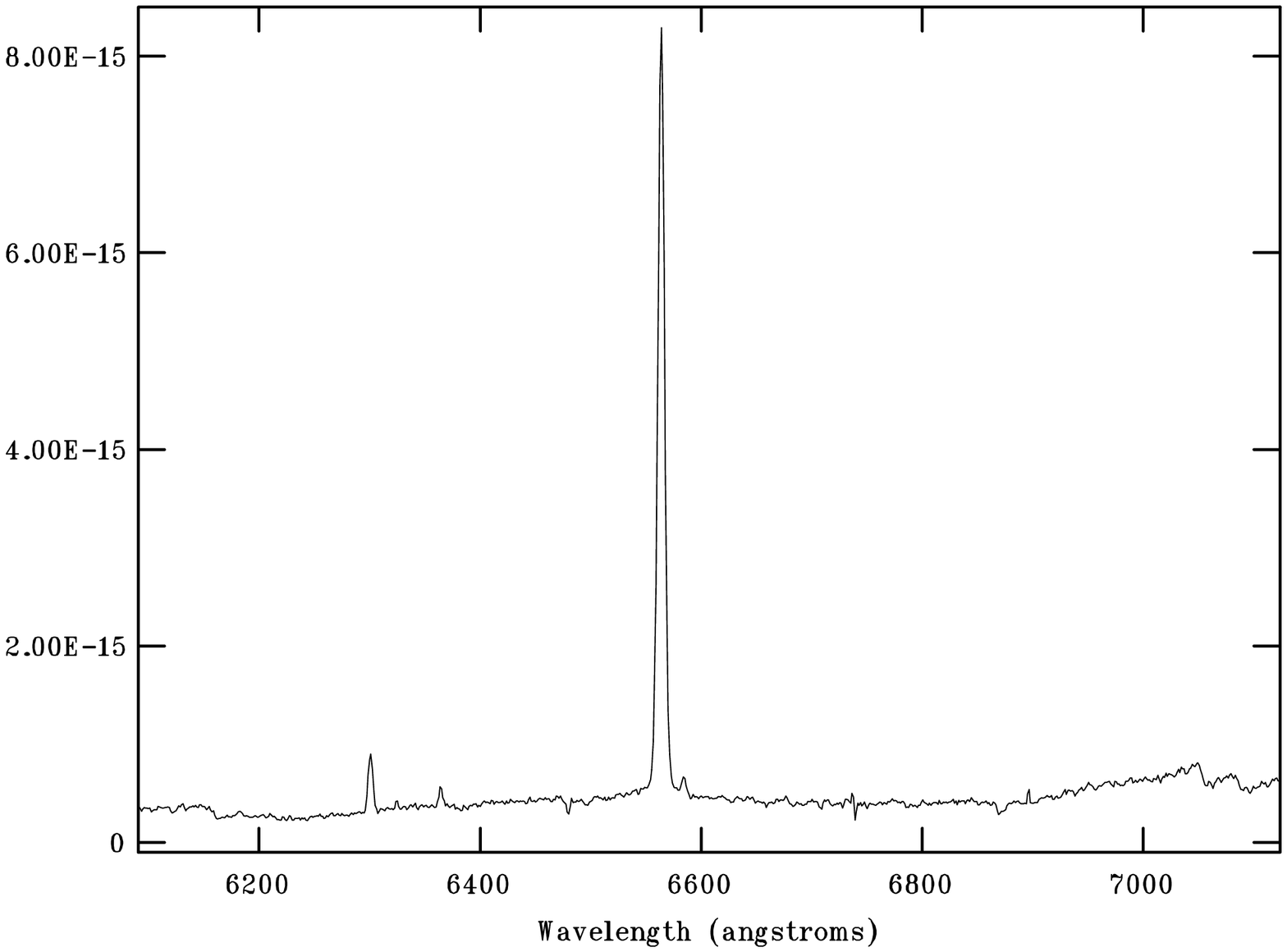}
\includegraphics[angle=90,scale=0.5]{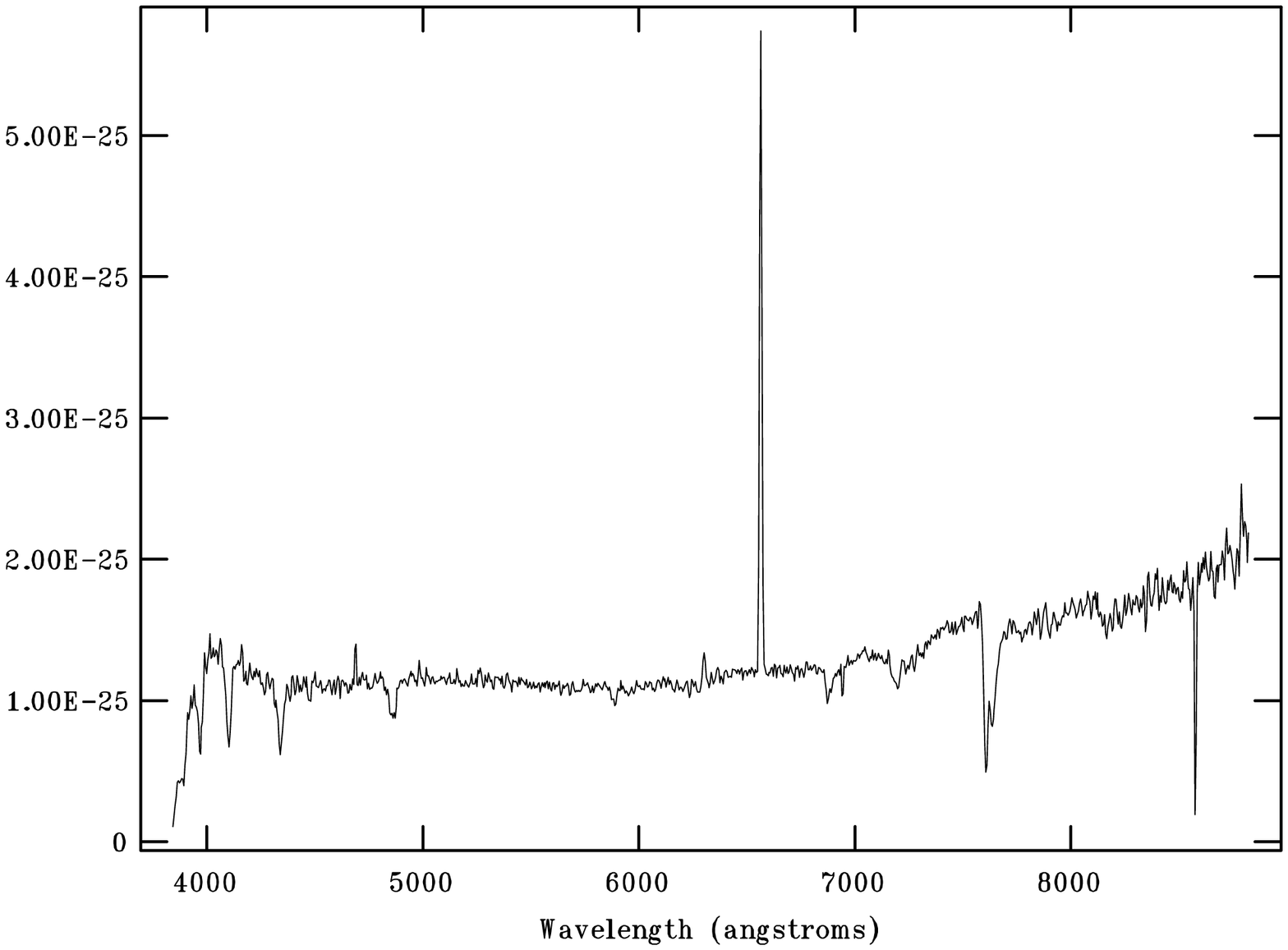}
\caption{Example spectra (top) for stars 20 (bottom) and 53.  Star 20 is a CTTS and shows a veiled 
continuum with strong H$\alpha$ and [\ion{O}{1}], 6300 and 6363~\AA\ emission lines.  Star 53 is an 
HAeBe star and shows the H$\alpha$ in emission but the other Balmer lines in absorption.}
\label{fig:spectra}
\end{figure}

Tables~3, 4, and 5 list, respectively, the CTTSs (plus some CTTS candidates), HAeBe stars, and 
non-PMS sources identified from spectroscopic observations.  In Table~3, stars 1--31 CTTSs are 
in Orion and 32--40 CTTSs are in Lacerta.  In Table~4, stars 41--48 are HAeBe stars in the Orion 
region, whereas the others are in the Lacerta region.  We derive the H$\alpha$, [\ion{O}{1}], and 
[\ion{S}{2}] equivalent widths of the CTTSs.  Some of the CTTSs listed in Table~3 do not show lithium 
absorption, but exhibit other CTTS characteristics, such as the H$\alpha$, \ion{Ca}{2}, and/or 
forbidden [\ion{O}{1}] and [\ion{S}{2}] emission line(s) in their spectra.  Since most of these 
spectra show veiling, their Li absorption line might be veiled by continuum radiation.  Thus 
they are included in the CTTS sample (Table~3) even though the Li line is not readily 
discernible.  CTTSs without a Li absorption line are not unusual; recently \citet{whi05} also 
found a lithium-depleted CTTS, St\,34, in the Taurus-Auriga T association.  
Figure~\ref{fig:spectra} presents an example of the spectra of a CTTS and a HAeBe star.  No PMS 
stars were found in any of the two control fields; most of the sources there are either carbon 
stars or M giants.

\begin{figure}
\epsscale{1.0}
\includegraphics[angle=270,scale=0.7]{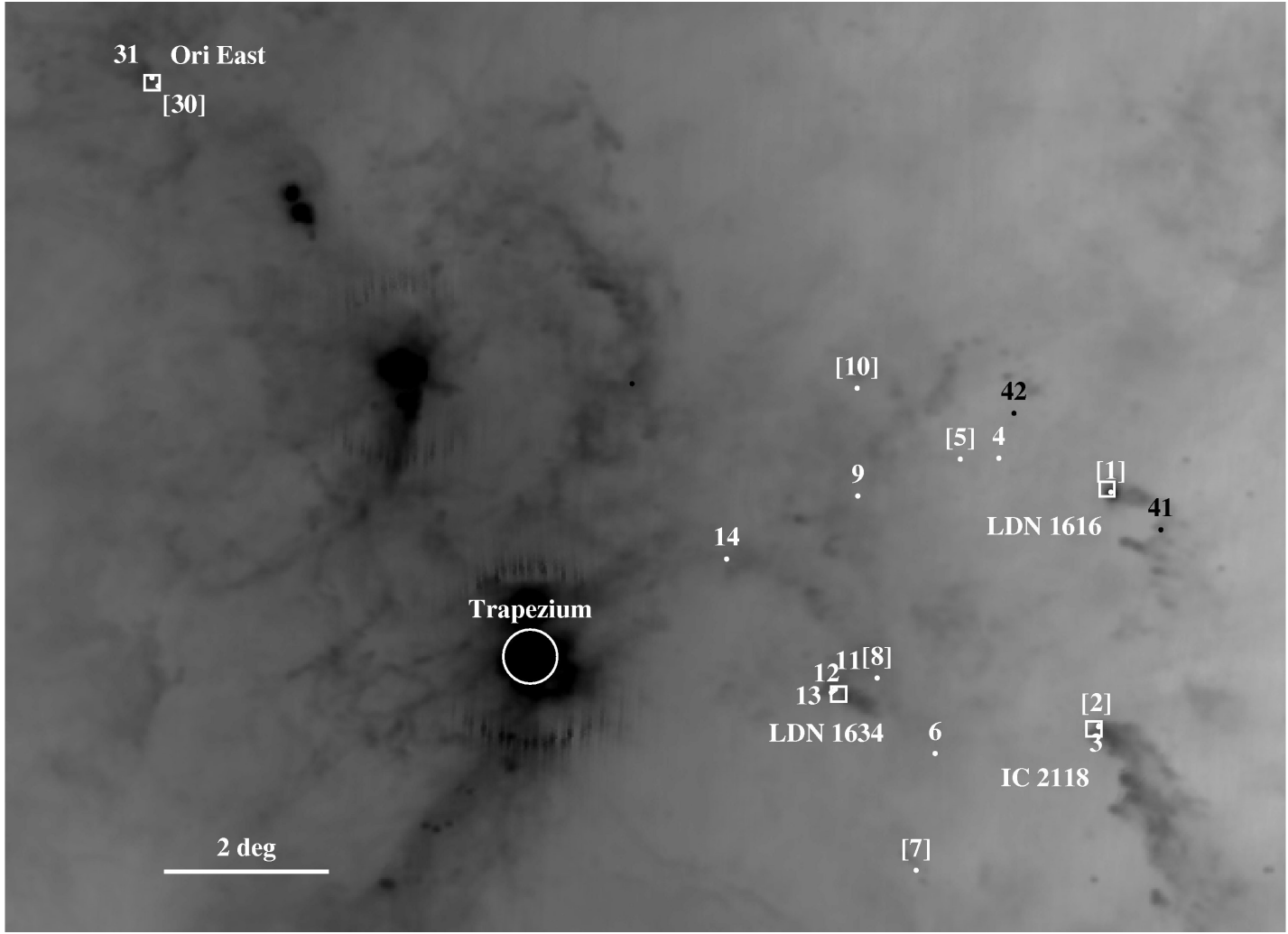}
\caption{$IRAS$ 100~$\micron$ image of the Trapezium region.  The dots indicate CTTSs (white) and 
HAeBe stars (black), labeled with the identification numbers from Tables~3 and 4.  PMS stars with 
forbidden line(s) are bracketed.  The boxes mark the fields of the H$\alpha$ images shown in
(Fig.~\ref{fig:lot1}).  East is to the left and north to the top.}
\label{fig:oriob1}
\end{figure}

\begin{figure}
\epsscale{1.0}
\includegraphics[angle=270,scale=0.5]{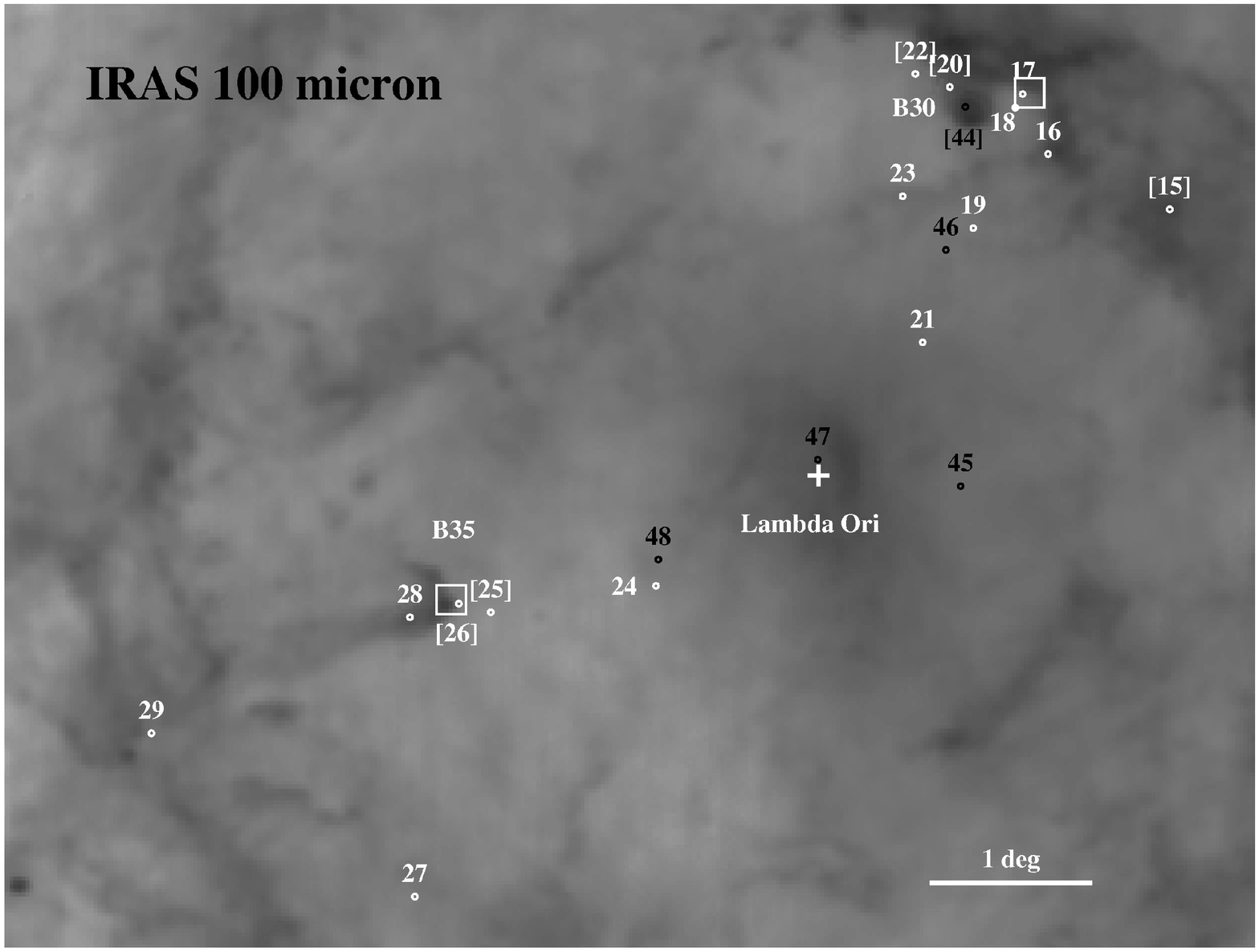}
\includegraphics[angle=270,scale=0.5]{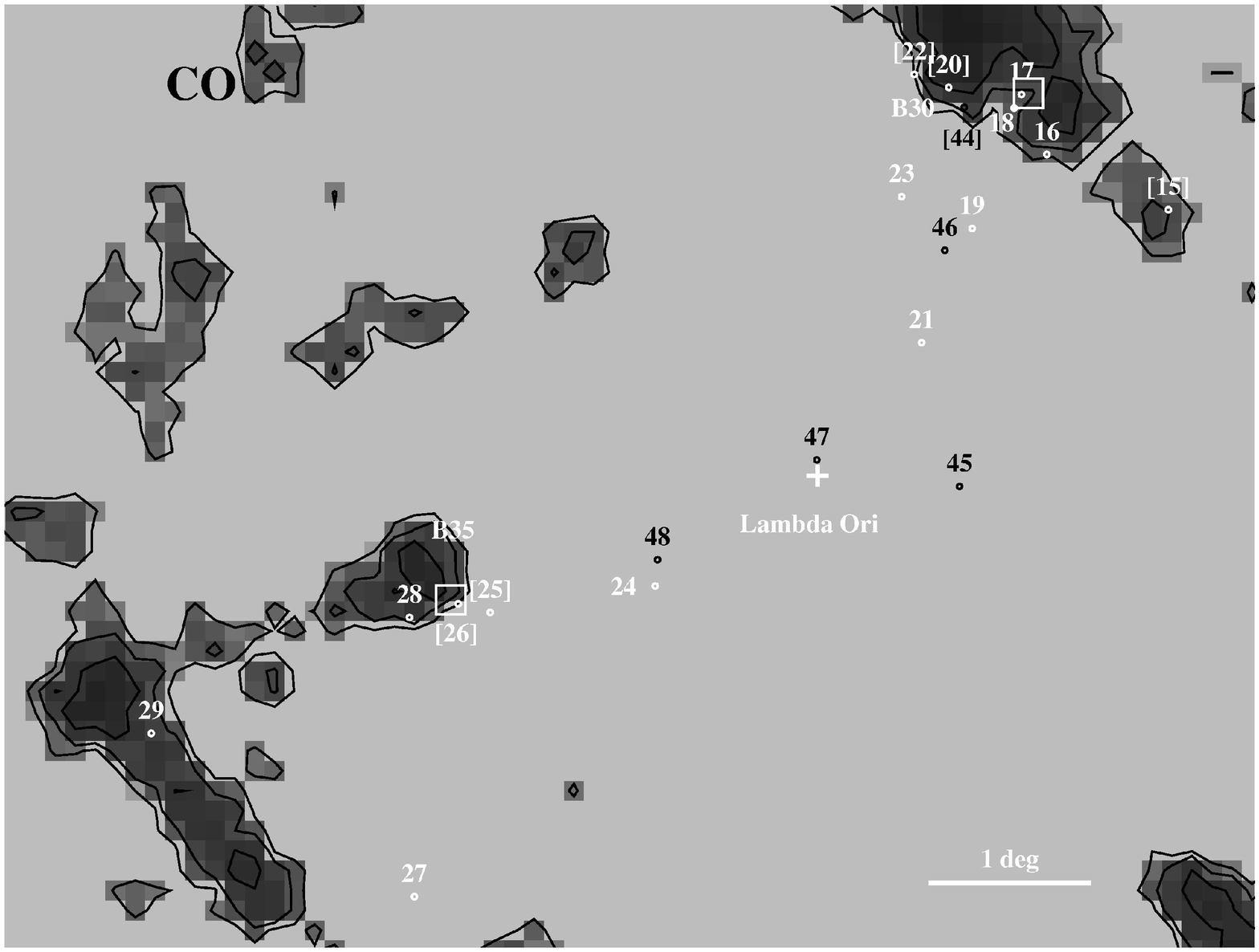}
\caption{$IRAS$ 100~$\micron$ and CO images of the $\lambda$ Ori region.  The symbols are the same 
as in Figure~\ref{fig:oriob1}.  The distribution of PMS stars extends from $\lambda$ Ori to B\,30 
and B\,35.  East is to the left and north to the top.}
\label{fig:lamori}
\end{figure}

\begin{figure}
\epsscale{1.0}
\includegraphics[angle=0,scale=0.38]{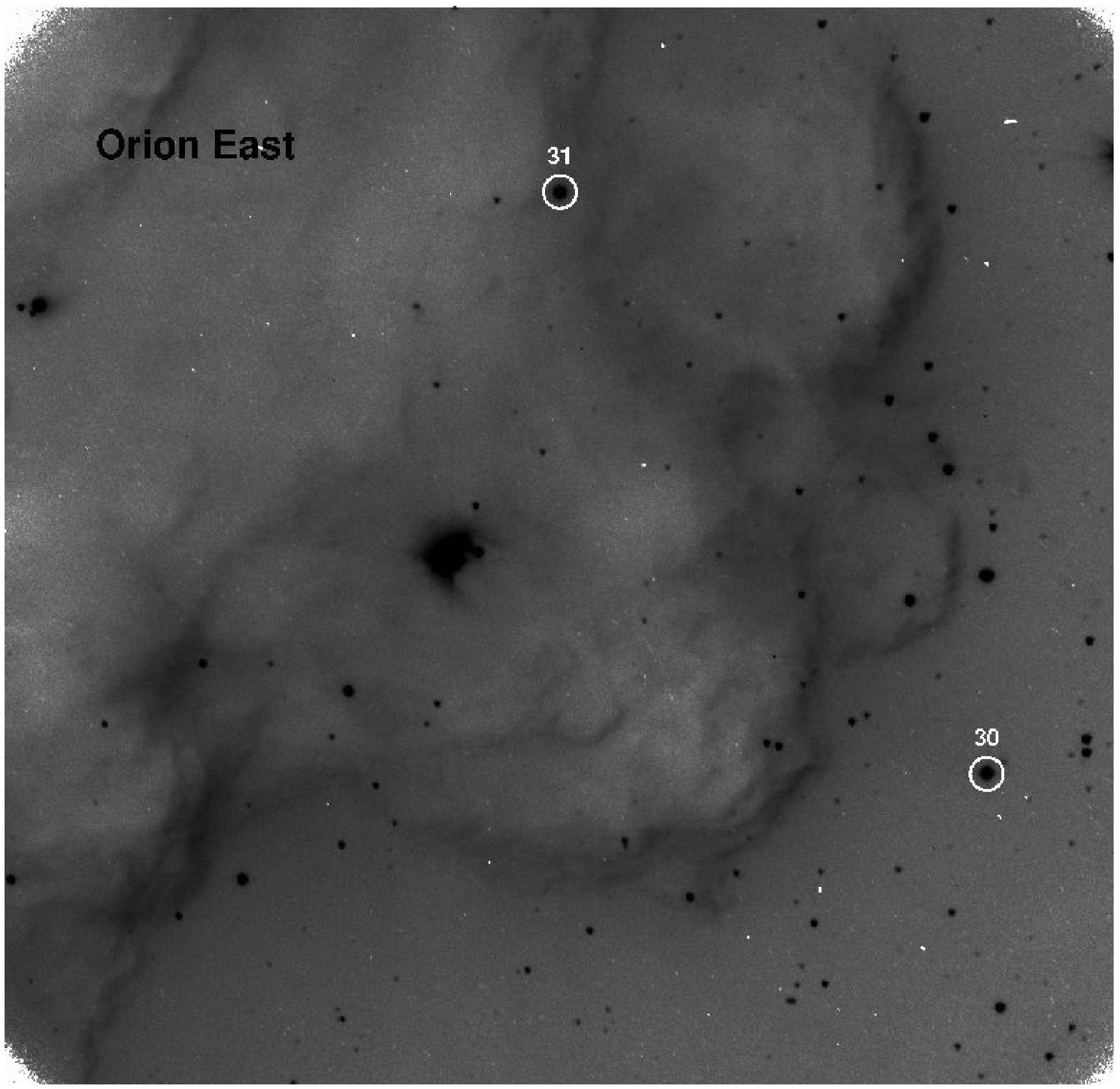}
\includegraphics[angle=0,scale=0.38]{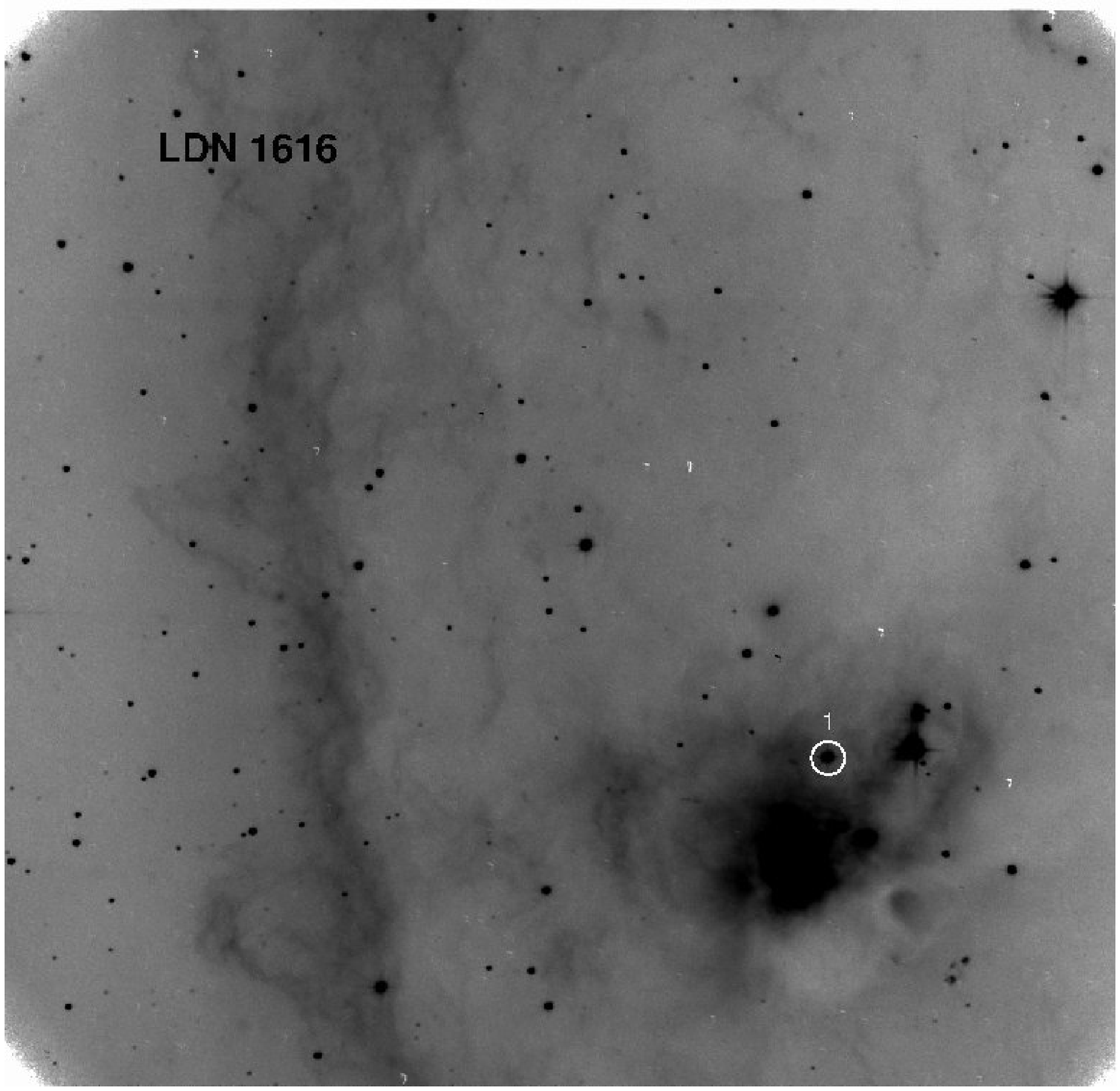}
\includegraphics[angle=0,scale=0.38]{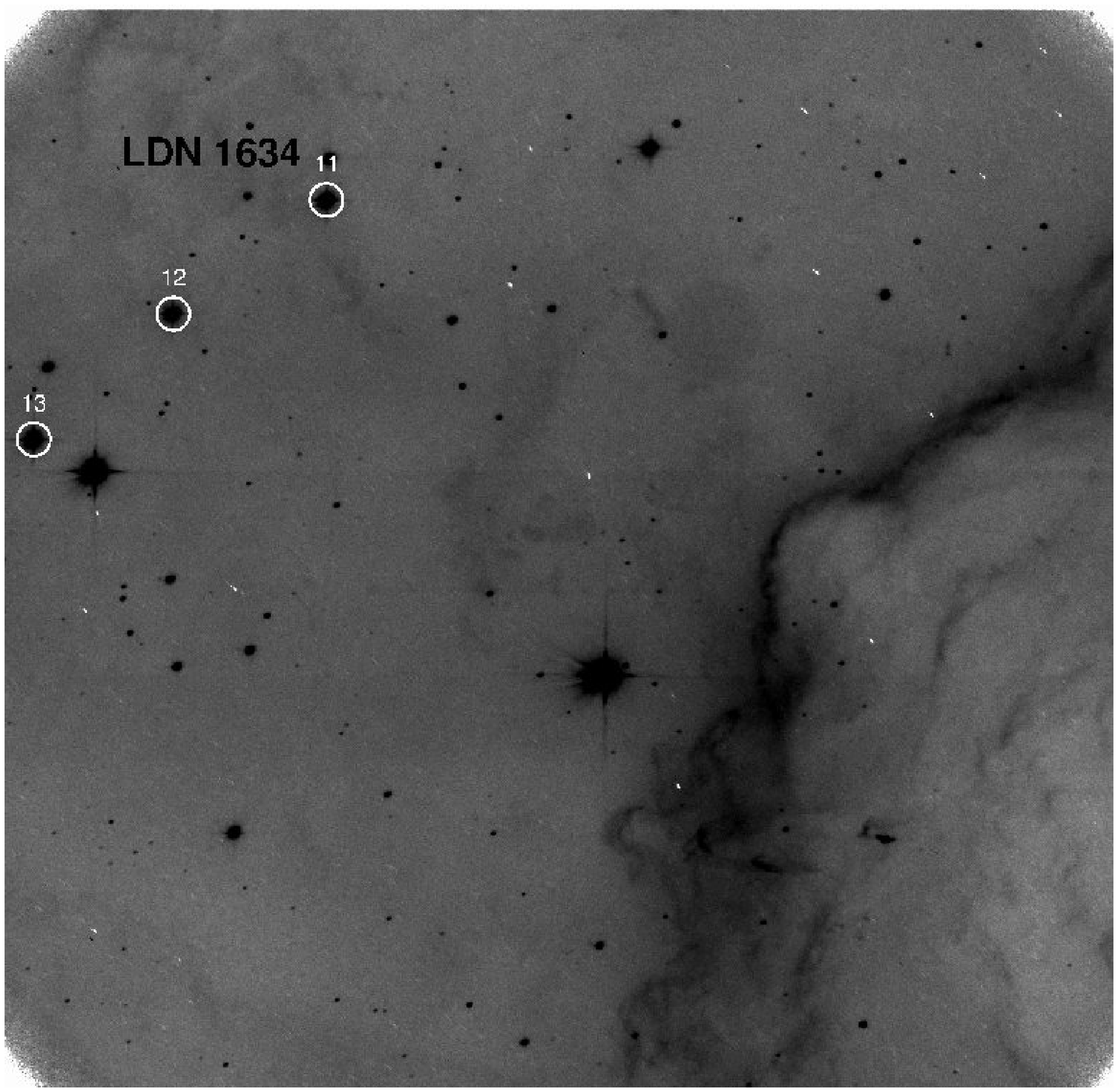}
\includegraphics[angle=0,scale=0.38]{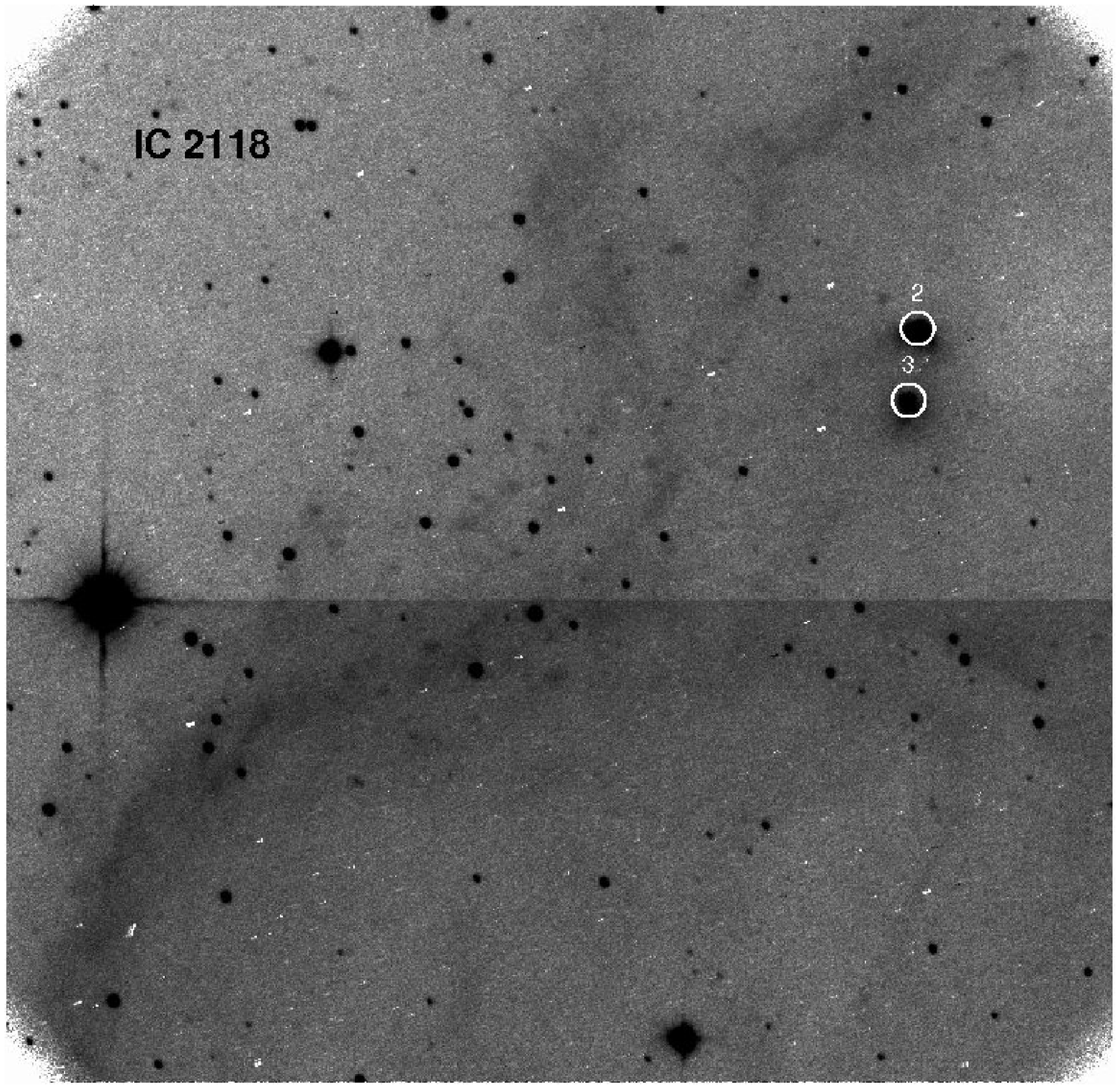}
\includegraphics[angle=0,scale=0.38]{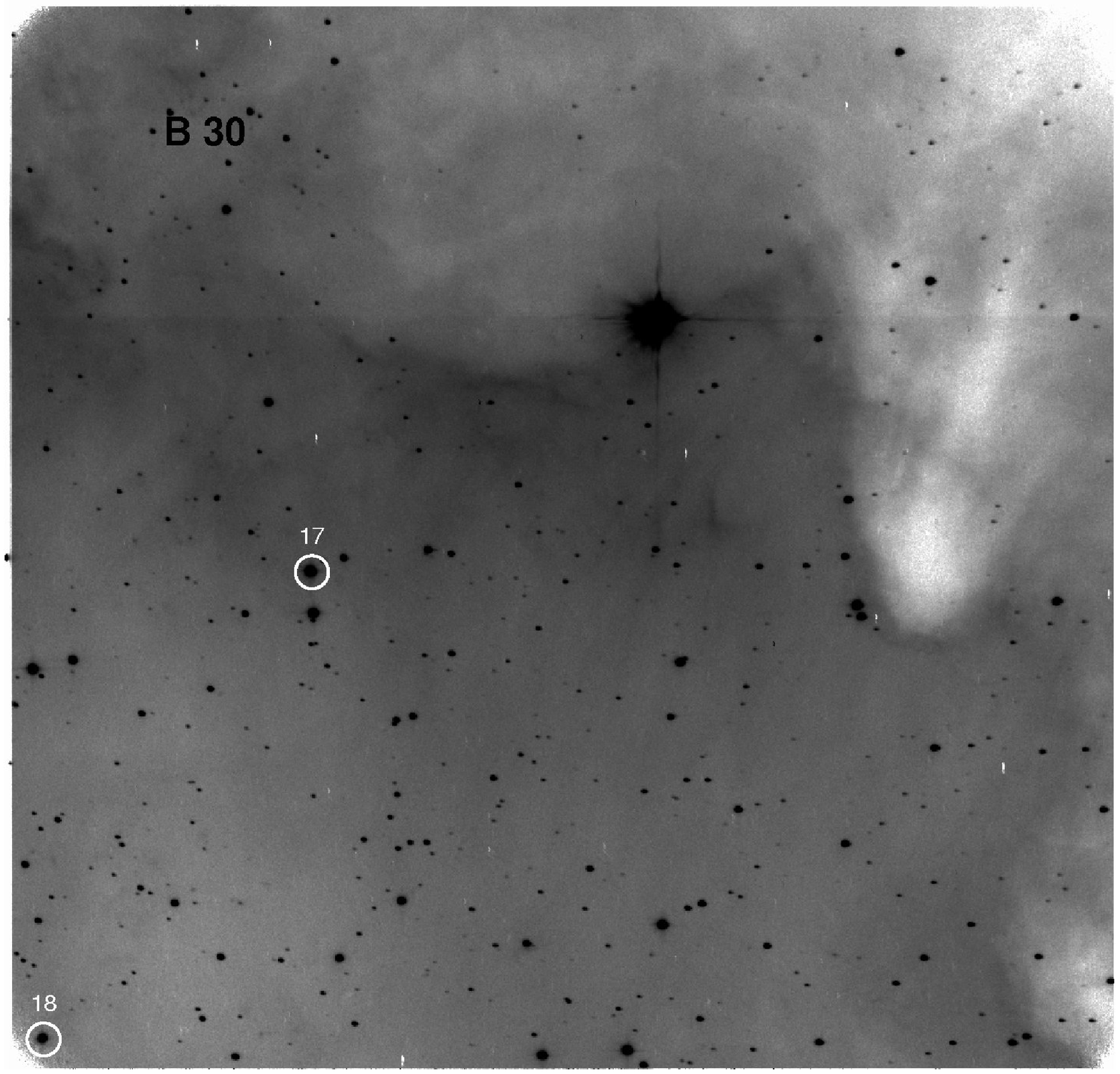}
\includegraphics[angle=0,scale=0.38]{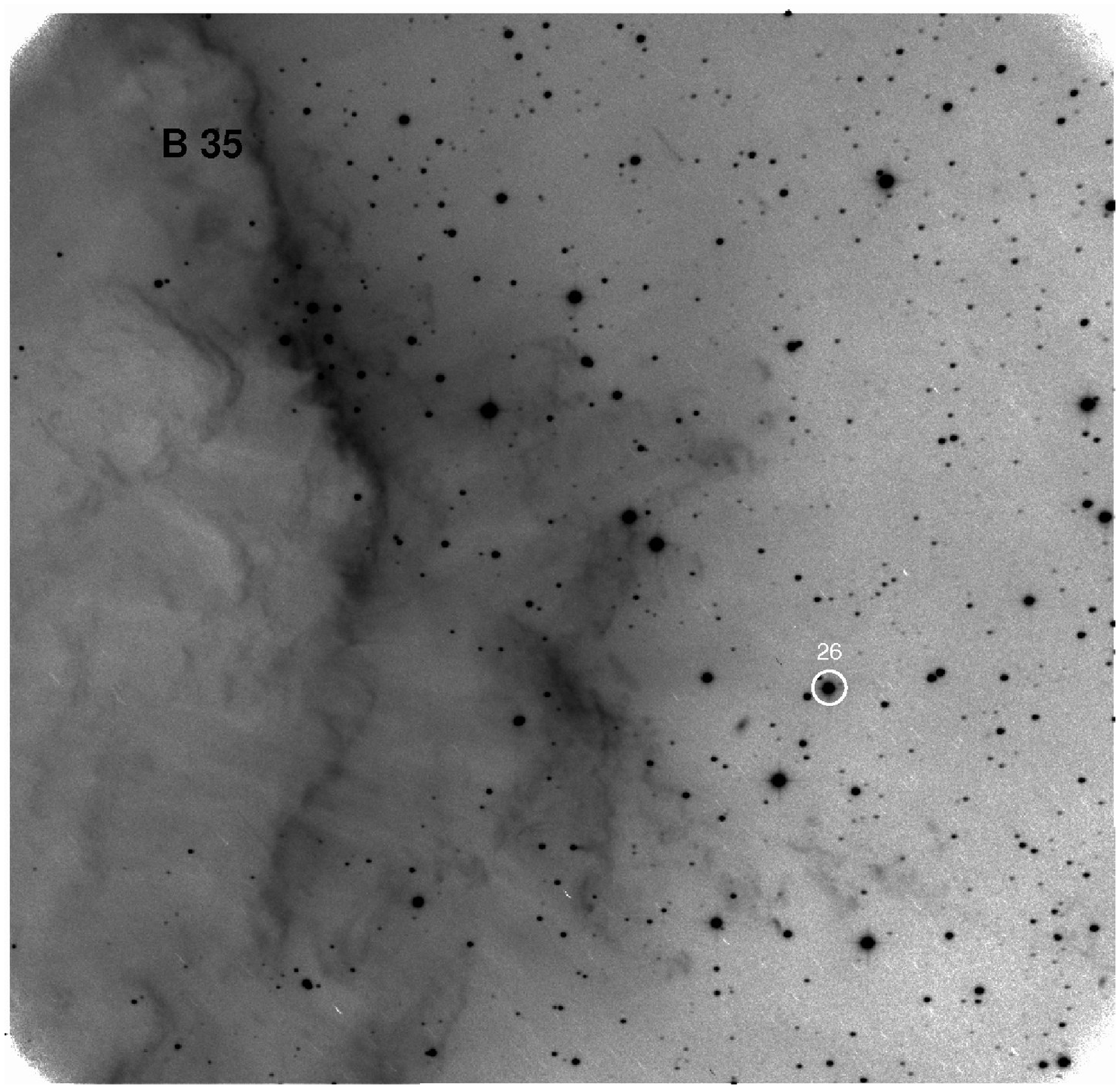}

\caption{H$\alpha$ images of the Ori OB1 BRCs.  The stars in Table~3 are marked.  East is to the 
left and north to the top.  The field of view of each image is $\sim 11\arcmin$.}
\label{fig:lot1}
\end{figure}

Figure~\ref{fig:oriob1} and Figure~\ref{fig:lamori} show, respectively, the Trapezium and the 
$\lambda$\,Ori regions in Orion, with the CTTSs (stars 1--31 in Table~3) and HAeBe stars (stars 
41--48 in Table~4) being marked.  The boxes mark the fields of the H$\alpha$ images presented in 
Fig.~\ref{fig:lot1}.  It is clear that the BRCs are outlined by the H$\alpha$ emission, and that 
some PMS stars are spatially close to the I-fronts.

\begin{figure}
\epsscale{1.0}
\includegraphics[angle=0,scale=0.4]{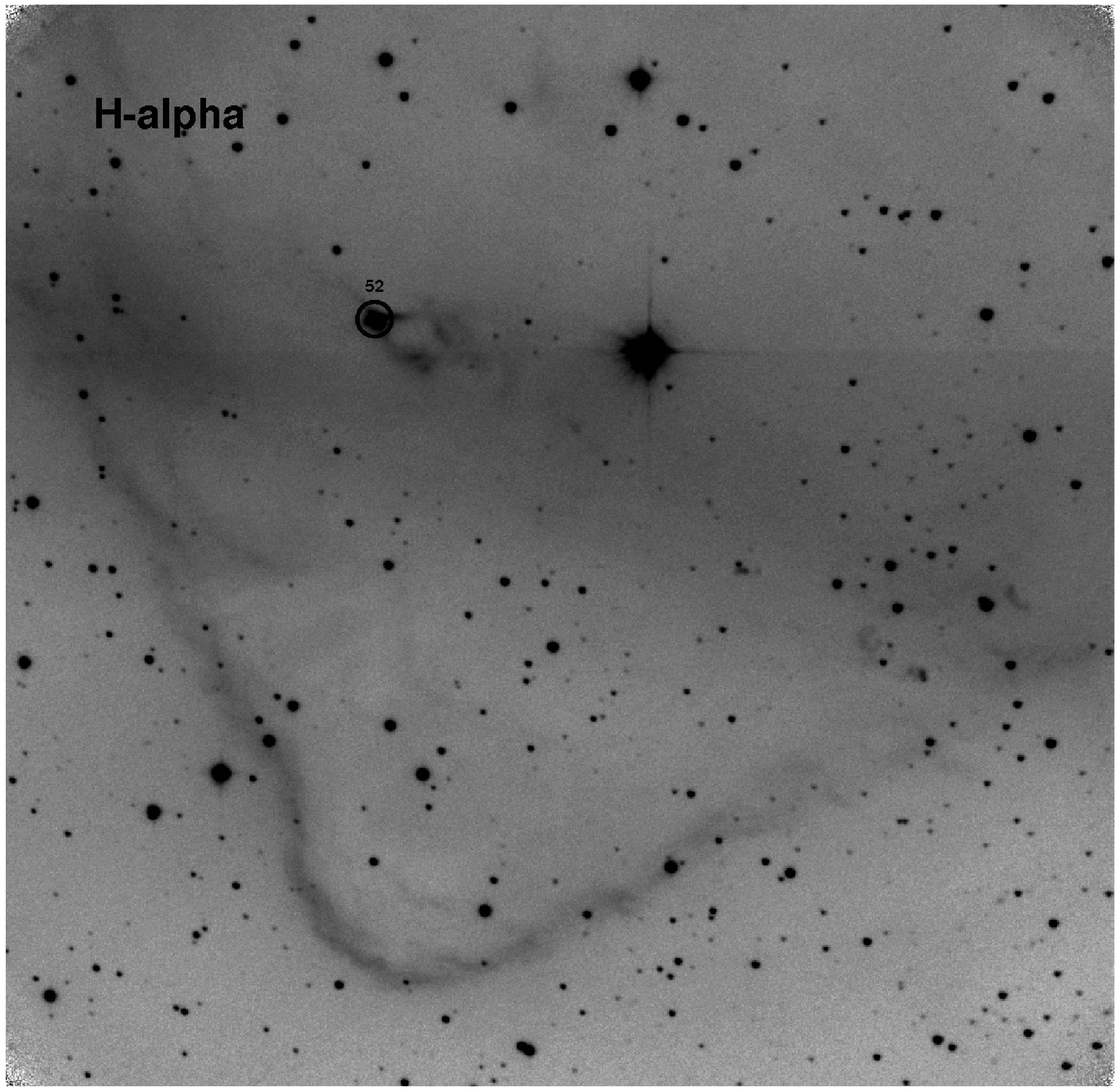}
\includegraphics[angle=0,scale=0.4]{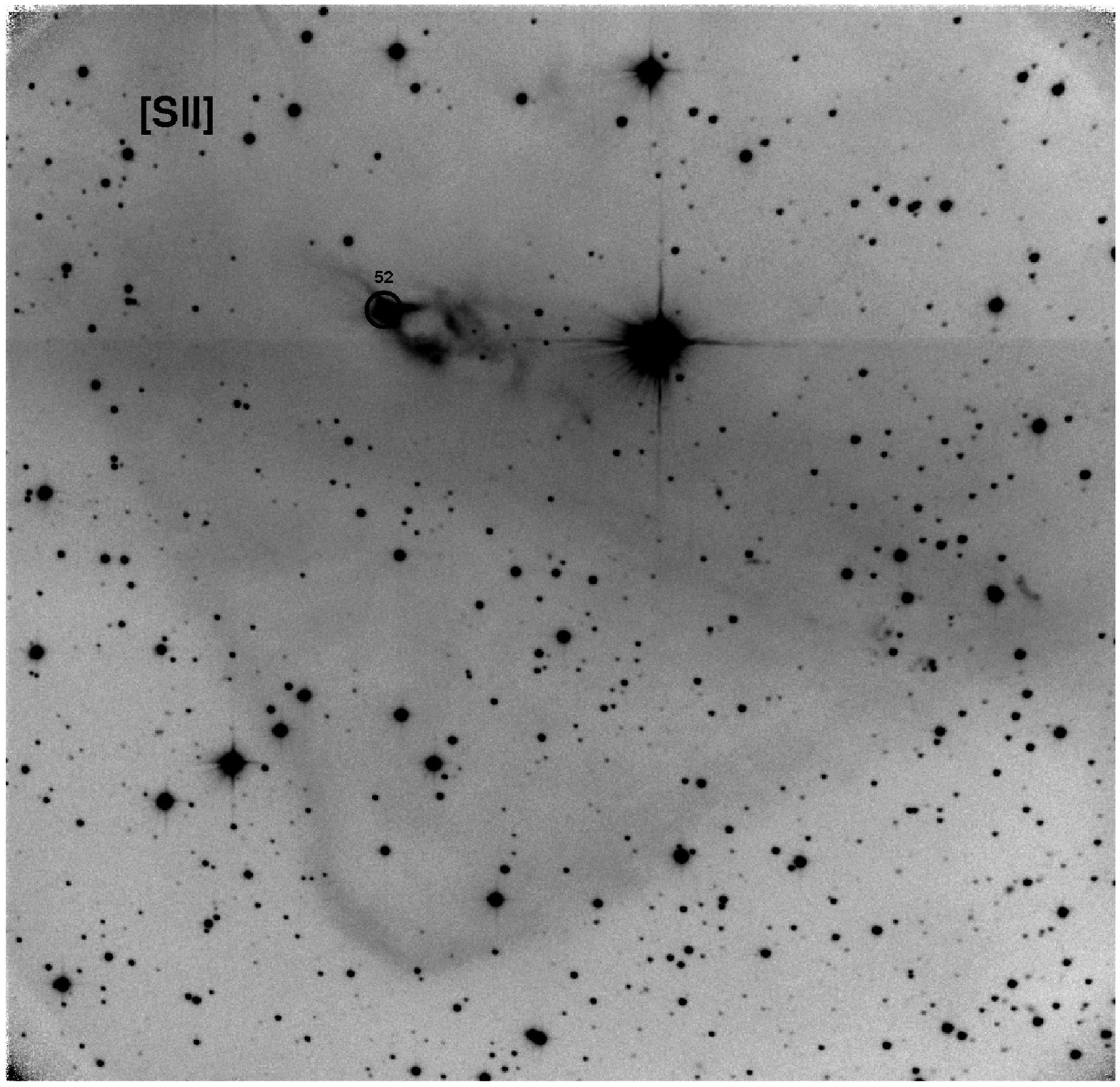}
\caption{H$\alpha$ and [\ion{S}{2}] images of LBN\,437.  Star 52 is associated with the nebulosity 
HH\,398.  East 
is to the left and north to the top.  The field of view of each image is $\sim 11\arcmin$}
\label{fig:lot2}
\end{figure}

Figure~\ref{fig:lac} displays the $IRAS$ 100~$\micron$, H$\alpha$ and CO emission maps of the 
Lac\,OB1 association.  The PMS stars in Table~3 and Table~4 are again marked.  The box indicates 
the LBN\,437 region shown in Figure~\ref{fig:lot2}.  LBN\,437 is a comet-shaped BRC 
\citep{ola94}.  The HAeBe star V375\,Lac (star~52 in Table~4) associated with this cloud is 
believed to be the exciting source of the parsec-scale Herbig-Haro outflow HH\,398 \citep{mcg04}.

\begin{figure}
\epsscale{1.0}
\includegraphics[angle=90,scale=0.6]{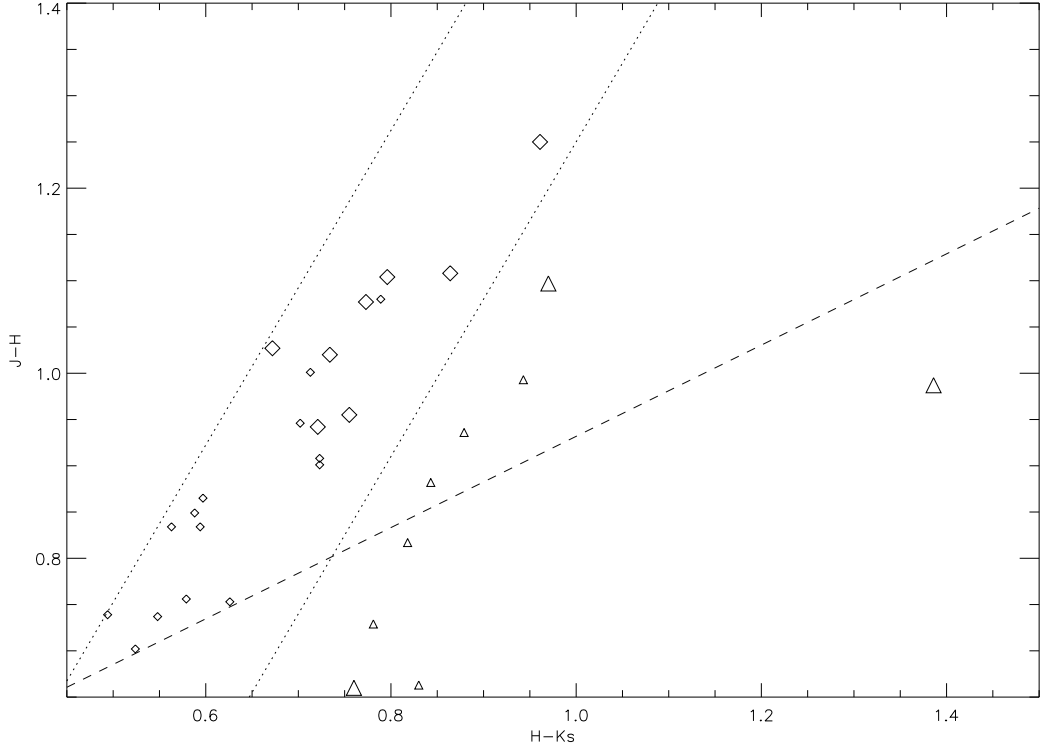}
\caption{2MASS color-color diagram of the CTTSs (diamonds) and HAeBe stars (triangles) from Tables~3 
and 4.  Here we only present the CTTSs with a Li absorption line detected in their spectra.  CTTSs 
and HAeBe stars with forbidden lines are indicated by larger symbols.  PMS stars with forbidden 
line are redder than those without.  The dotted and dashed lines represent the reddening direction 
and the dereddened CTTS locus, respectively.  CTTSs and HAeBe stars are well separated by the line 
$(j\_m-h\_m)-1.7(h\_m-k\_m)+0.450=0$.}
\label{fig:colors}
\end{figure}

In Paper~I it was shown that CTTSs exhibiting continuous or veiled spectra with [\ion{O}{1}] and/or 
[\ion{S}{2}] forbidden lines, originating from jets or winds seen commonly in Class~I 
sources \citep{ken98}, tend to be redder, which is suggestive of a younger age, than those 
without.  A color-color diagram of the PMS stars in Tables~3 and 4 is plotted as Fig.~\ref{fig:colors}; 
the results agree with our previous work.  This correlation extends to HAeBe stars, in that 
HAeBe stars with forbidden line(s) are mostly located on the upper right of the 2MASS color-color 
diagram.  As an alternative to being younger, a CTTS with forbidden lines could be the result of 
reduced photoevaporation of the circumstellar disk, e.g., by being away from a luminous star or 
shielded by a molecular cloud.  Only 14 of the 40 CTTSs and 4 of the 13 HAeBe stars in our sample 
show forbidden line(s) in their spectra.  In other words, about one third of the PMS stars with 
strong infrared excess exhibit forbidden line(s).  Typical CTTS ages are a few Myr 
\citep{ken95}, with those with forbidden lines representing an even younger sample, probably no 
more than a couple Myr old.  The [\ion{S}{2}] line is only present in Star 2, a CTTS with a strong 
infrared excess and strong [\ion{O}{1}] (equivalent width $>$ 10.5 \AA).  In our sample of CTTSs 
there is no correlation between the H$\alpha$ equivalent widths and the presence of forbidden 
lines, or between the H$\alpha$ equivalent widths and the 2MASS colors.

The success rate of spectroscopic confirmation of CTTS and HAeBe candidates is extremely high for 
$\lambda$ Ori, Ori\,OB1 or Lac\,OB1.  Candidates closely associated with star-forming regions all 
turned out to be bona fide young stars with essentially no exception, whereas the regions away 
from molecular clouds are mostly populated by evolved stars (e.g., carbon stars or M giants).  The 
2MASS database enables us to effectively trace recent star formation on a large scale, without any 
a priori bias toward prominent \ion{H}{2} or reflection nebulae which are obvious targets to search 
for young stellar objects.  For example, stars 35--38 in our sample are confirmed to be young 
stars.  They are located away from prominent nebulosity, so it might otherwise be difficult to 
recognize them as young stars in a targeted survey.

\section{STAR FORMATION IN THE ORI\,OB1 AND LAC\,OB1 ASSOCIATIONS}

Star formation triggered by the RDI mechanism has several characteristics which can be diagnosed 
observationally: (1)~The remnant cloud is extended toward, or pointing to, the massive stars. 
(2)~The young stellar groupings in the region are roughly lined up between the remnant clouds and 
the luminous star.  (3)~Stars closer to the cloud, which have formed later in the sequence, are 
younger in age, with the youngest stars being in the interacting region, i.e., along the bright 
rim of the cloud.  (4)~No young stars exist far behind the BRC.  In particular, (3) and (4) are in 
distinct contrast to the case of spontaneous star formation, which conceivably would not have left 
such distinguishing temporal and positional signposts.  In Table~6 we summarize the different 
outcomes of the triggered versus spontaneous star formation processes.

In Paper~I, we presented evidence supporting the induced star formation in six Orion BRCs, namely 
B\,30, B\,35, Ori\,East, IC\,2118, LDN\,1616, and LDN\,1634.  Here, we present further 
spectroscopic observations of the Orion sources, classified as young star "candidates" in Paper~I 
and extend our sample to include the Lac\,OB1 region.  Combined with the earlier Ori\,OB1 results, this 
reinforces the links between massive stars, BRCs, and the formation of low-mass stars.  
Furthermore, our young star sample now contains not only CTTSs, but also young intermediate-mass 
stars, rendering a more comprehensive understanding of the origin of stellar masses in an OB 
association.

\subsection{STAR-FORMING ACTIVITIES IN THE ORION REGION}

\subsubsection{IC\,2118, LDN\,1616, LDN\,1634, AND ORI\,EAST}

IC\,2118, LDN\,1616, and LDN\,1634 are three isolated BRCs around the Trapezium to the west of the 
Orion\,A.  Another BRC, Ori\,East, can be found to the north-east of the Trapezium.  All these 
BRCs point roughly to the Trapezium (Fig.~\ref{fig:oriob1}), indicative of the Trapezium and/or 
the Orion-Eridanus superbubble being the shaping source of these BRCs 
\citep{alc04,sta02,kun01,kun04}.

In this region most CTTSs with forbidden line(s), i.e., those of younger ages, are spatially close 
to the BRCs, e.g., stars 1, 2, and 30 in relation to LDN\,1616, IC\,2118, and Ori\,East, 
respectively.  Star~7 is also associated with a remnant molecular cloud \citep{ogu98} (their cloud~6).

\subsubsection{B\,30 AND B\,35}

B\,30 and B\,35 are two BRCs associated with an \ion{H}{2} region excited by the O8~III star 
$\lambda$ Ori and surrounded by a ring-shaped molecular cloud \citep{lan00}.  \citet{due82} find 
some 80 H$\alpha$ stars in the $\lambda$ Ori region, most of which are distributed as a barlike 
structure extending from either side of $\lambda$\,Ori to B\,30 and to B\,35.  
\citet{dol99,dol01,dol02} present photometric and spectroscopic studies of the young stellar 
population in the $\lambda$ Ori region.  They suggested the ring-shaped molecular cloud to be 
caused by a supernova explosion that terminated recent star formation in the vicinity.

It is likely that $\lambda$ Ori is the triggering source responsible for the star formation in B\,30 
and B\,35.  It is found that photoevaporative flows \citep{hes96} stream out of the surfaces 
of them; this is a demonstration of the interaction between a massive star and a molecular cloud 
(Fig.~\ref{fig:lamori}).  Here again we see that stars with forbidden lines, i.e., stars 
20, 22, and 44 in relation to B\,30, and stars 25 and 26 to B\,35, are all physically close to a 
BRC.  

\subsection{STAR FORMATION HISTORY IN ORI\,OB1 AND THE $\lambda$ ORI REGION}

It is suggested that star formation is triggered by the O stars and/or by the superbubbles in LDN\,1616 
\citep{alc04,sta02} and in IC\,2118 \citep{kun01,kun04}.  All BRCs in Ori\,OB1 which show evidence 
of star formation being triggered by nearby massive stars are found associated with strong $IRAS$ 
100~$\micron$ and H$\alpha$ emission (Fig.~\ref{fig:lot1}).  In every case a sequential 
process---that PMS stars closer to the triggering stars are older than those closer to the BRCs---can 
be clearly witnessed (Paper~I).

The same phenomena are also seen near $\lambda$ Ori.  Initially, the B\,30 and B\,35 clouds might 
have extended toward $\lambda$ Ori, perhaps forming a barlike structure.   The I-fronts from 
$\lambda$ Ori then propagated through the clouds, prompting star formation on both sides, thereby 
resulting in the lining up of the PMS stars, in an age sequence, between $\lambda$ Ori and the 
B\,30 and B\,35 clouds.


In Paper~I we show that the CTTSs that are spatially close to BRCs are among the brightest,  
just revealing themselves on the birthline and beginning to descend down the Hayashi tracks.  We also 
find no young stars far behind the I-fronts, i.e., embedded in the BRCs.  These photoevaporated 
clouds typically have low extinction so that any PMS stars cannot have escaped the 2MASS 
detection.  In both Ori\,OB1 and the $\lambda$ Ori region therefore, we see the predomination of 
triggered star formation, as evinced in the cloud morphology, star grouping orientation, and star 
formation sequence.

\subsection{STAR FORMATION ACTIVITIES IN LAC\,OB1}

The Lac\,OB1 association, at a distance of $\sim 360$~pc \citep{dez99}, is one of the nearest OB 
associations.  \citet{bla58} divides Lac\,OB1 into 2 subgroups, ``a" and ``b", on the basis of 
stellar proper motions and radial velocities.  The entire Lac\,OB1 covers the region of the sky 
from $90\arcdeg < \ell < 110\arcdeg$ and $-5\arcdeg < b < -25\arcdeg$ \citep{dez99}.  Lac\,OB1b 
occupies an area with a radius of $\sim 5\arcdeg$ centered around $(\ell, b)=(97\arcdeg.0, 
-15\arcdeg.5)$ and Lac\,OB1a occupies the remaining area.  The Lac\,OB1b harbors the only O star 
(O9 V), 10\,Lac, in the Lac\,OB1 association.  Our study discusses two regions in Lac\,OB1 known to 
have current star-forming activities, LBN\,437 and GAL\,110-13, a BRC and a comet-shaped 
cloud, respectively.  

\subsubsection{LBN\,437}

\begin{figure}
\epsscale{1.0}
\includegraphics[angle=180,scale=0.38]{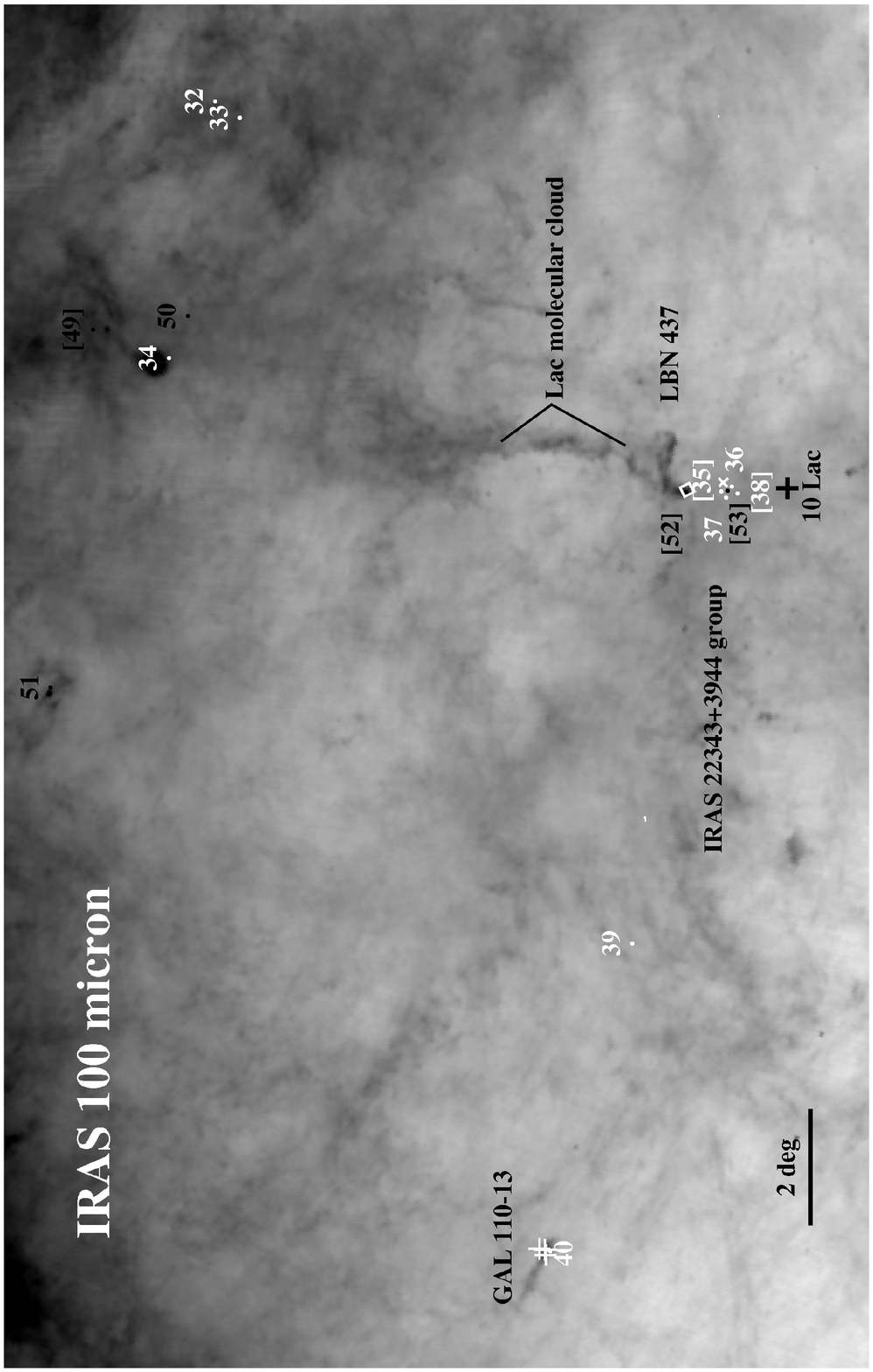}
\includegraphics[angle=180,scale=0.38]{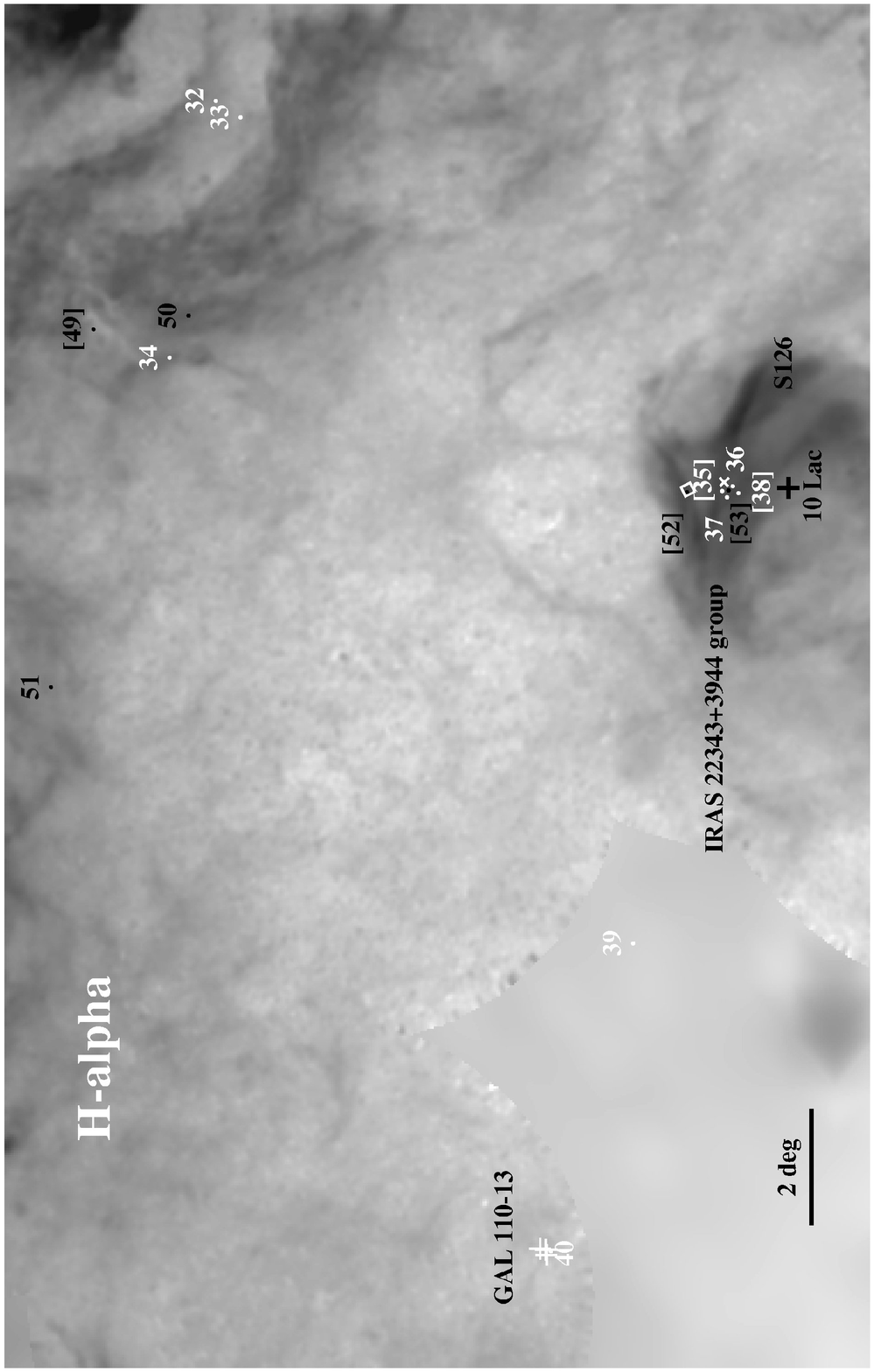}
\includegraphics[angle=180,scale=0.38]{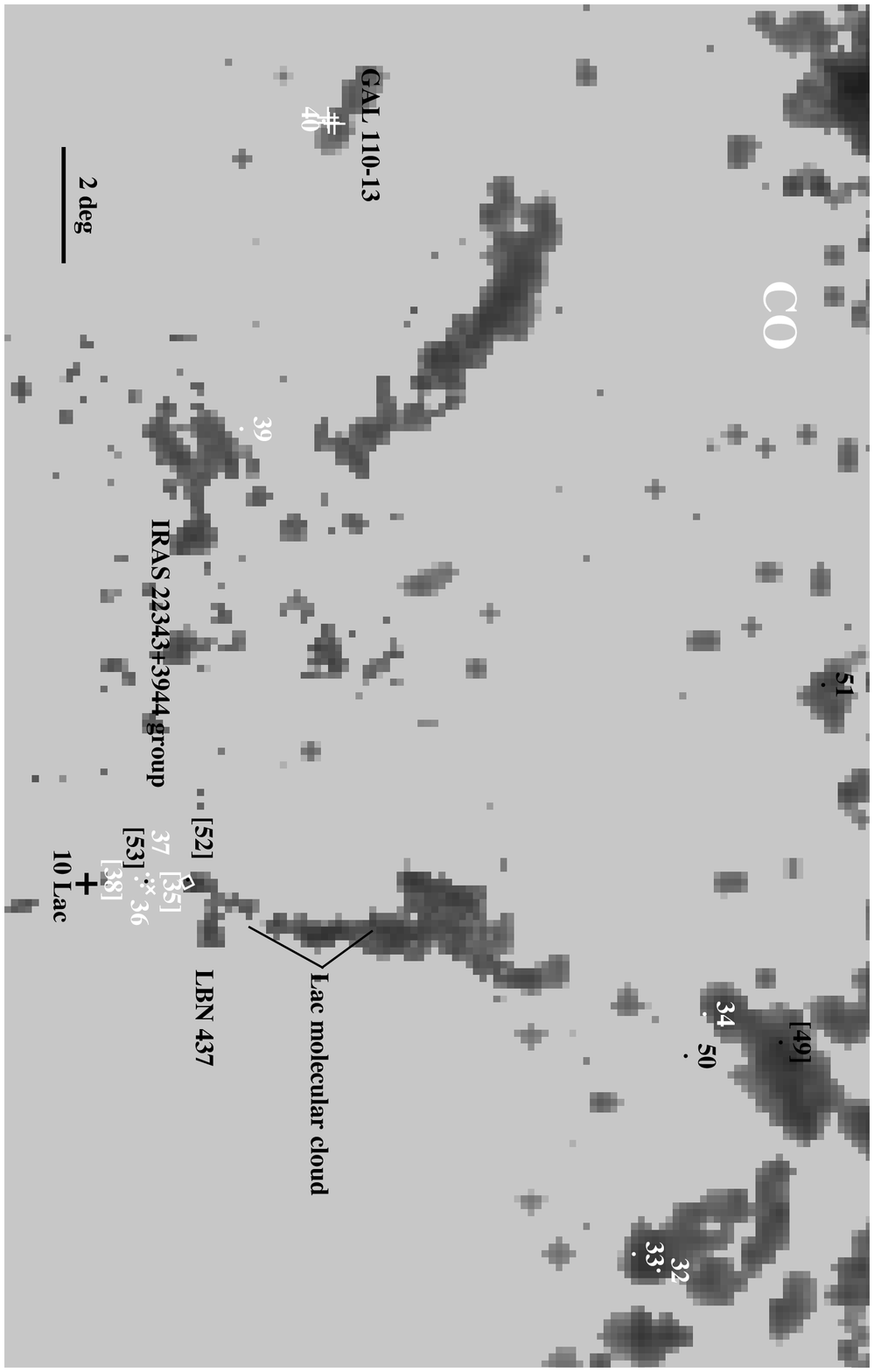}
\caption{$IRAS$ 100~$\micron$, H$\alpha$ and CO images of the Lac\,OB1 association using the 
orientation of the Galactic coordinates, i.e., north is to the top and the Galactic longitude 
increases to the left.  The white plus signs indicate the 3 late-B stars, HD\,222142, HD\,222086, 
and HD\,222046, in GAL\,110$-$13 and the cross marks indicate the CTTS candidate in the 
IRAS\,22343+3944 group, 2MASS\,J22354224+3959566.  The other symbols are the same as in 
Fig.~\ref{fig:oriob1}.}
\label{fig:lac}
\end{figure}

\begin{figure}
\epsscale{1.0}
\includegraphics[angle=270,scale=0.6]{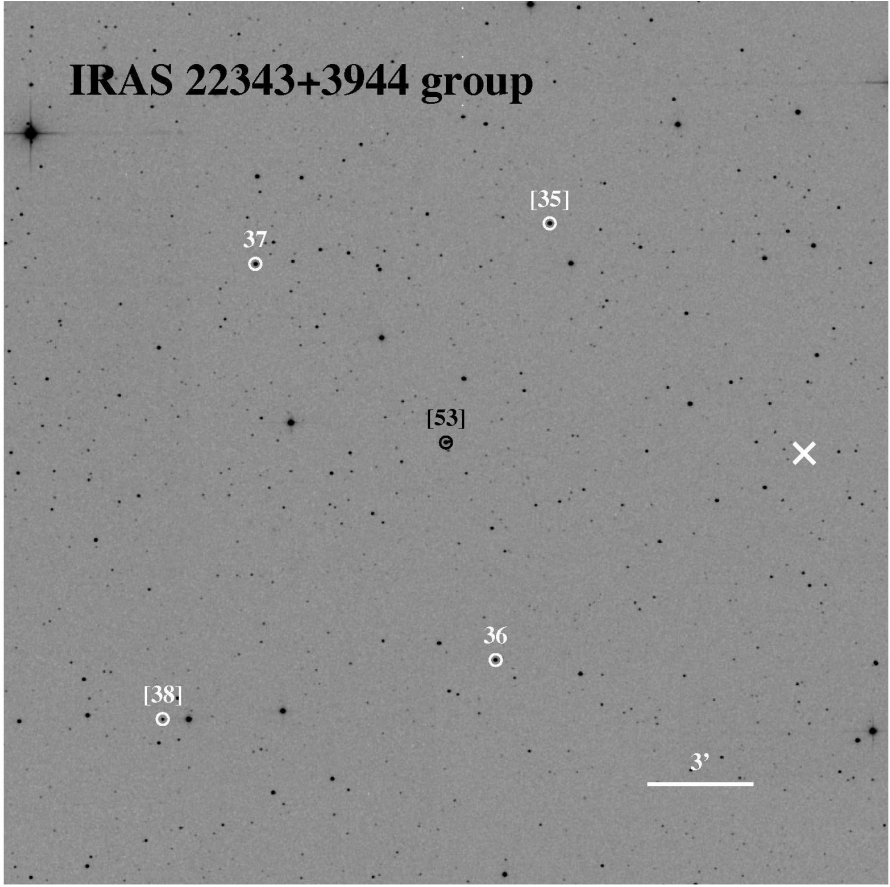}
\caption{2MASS K$_{s}$ image of the IRAS\,22343+3944 group.  CTTSs 35, 36, 37, and 38 from Table~3 
(in white) and HAeBe star 53 (in black) are labeled.  The cross indicates the CTTS candidate, 
2MASS\,J22354224$+$3959566.  The other symbols are the same as in Fig.~\ref{fig:oriob1}.  East is 
to the left and north to the top.}
\label{fig:group}
\end{figure}

LBN\,437 is at the edge of an elongated molecular cloud \citep{ola94} and on the border of the 
\ion{H}{2} region S\,126 excited by the nearby O star, 10\,Lac.  Hereafter we call this elongated 
molecular cloud the ``Lac molecular cloud" (Fig.~\ref{fig:lac}).  Between 10\,Lac and LBN\,437 
there is a small stellar group (Fig.~\ref{fig:group}) which includes 5 CTTSs (stars 35--38 in 
Table~3, plus the CTTS candidate 2MASS\,J22354224$+$3959566, for which we do not have 
spectroscopic observations) and one HAeBe star (star~53 in Table~4).  The HAeBe star is an $IRAS$ 
source, IRAS\,22343$+$3944.  We can identify IRAS\,22343$+$3944 as the counterpart of star~53, 
because this star shows a near-infrared excess and is located within the positional error for 
IRAS\,22343$+$3944.  Hereafter we refer to this 6-star system as the IRAS\,22343$+$3944 group 
(Fig.~\ref{fig:lac}).  The size of the IRAS\,22343$+$3944 group is about 24$\arcmin$, which 
corresponds to $\sim 2.5$~pc at 360~pc.

\subsubsection{GAL\,110$-$13}

\begin{figure}
\epsscale{1.0}
\includegraphics[angle=270,scale=0.6]{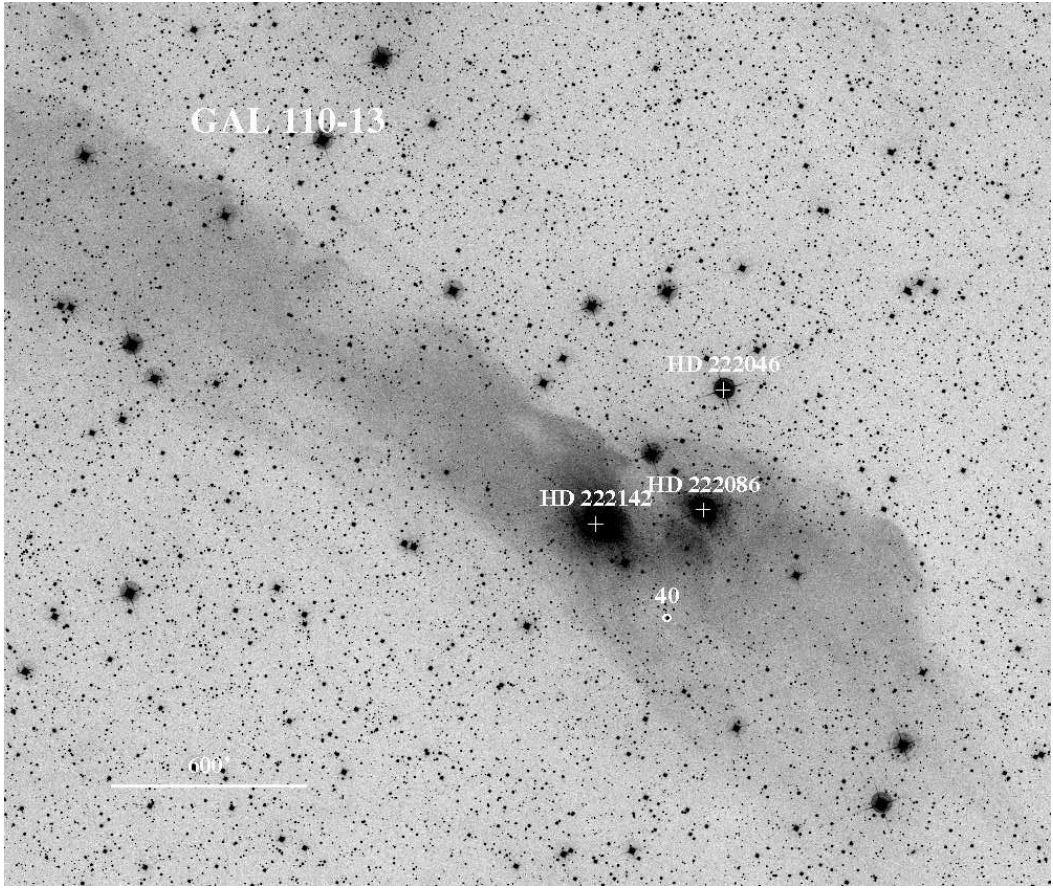}
\caption{DSS blue image of the comet-shaped cloud GAL\,110-13.  Star 40 (CTTS) and three late-B 
stars are marked.  The Galactic longitude and latitude increase to the left and to the top, respectively.}
\label{fig:gal}
\end{figure}

GAL\,110$-$13 is an isolated and elongated molecular cloud (Fig.~\ref{fig:gal}) at a 
distance of $\sim 440$~pc \citep{ode92}.  Its head-tail, comet-like shape suggests compression by 
ram-pressure, perhaps as a result of a recent cloud collision \citep{ode92}.  
Star formation takes place on the compressed side of GAL\,110$-$13, e.g., the location of  
the CTTS star 40 (BM\,And) and the nebula vdB\,158 reflecting light from the B9.5V star 
HD\,222142 \citep{mag03}.  In addition to HD\,222142 there are two other late B-type stars in the 
vicinity, HD\,222046 and HD\,222086.  All three B stars and star 40 share common proper motions 
\citep[data extracted from the Second U.S. Naval Observatory CCD Astrograph Catalog]{zac04}, which 
are consistent with those of the Lac\,OB1 groups \citep{esa97}, as summarized in Table~7.  
GAL\,110$-$13 is located near the border of the Lac\,OB1 association, at a distance not very 
different from that of Lac\,OB1.  GAL\,110$-$13 was not included as part of Lac\,OB1 by 
\citet{dez99}, but our analysis suggests that the cloud, together with the young stars associated 
with it, is likely part of Lac\,OB1a. 

GAL\,110$-$13 is elongated and roughly points toward 10\,Lac (see Fig.~\ref{fig:lac}).  This 
implies that Lac\,OB1b or 10\,Lac alone is responsible for shaping the cloud.  Either shock fronts 
from a supernova or ionization fronts from a massive star could have caused the shape of this 
cloud as well as the spatial distribution of young stars in GAL\,110$-$13.  In the supernova 
scenario a star in Lac\,OB1b more massive than 10\,Lac exploded, and, assuming that Lac\,OB1b and 
10\,Lac are at the same distance from us (i.e., 358~pc), it would take a few hundred thousand 
years for the supernova shock waves (at a speed of hundreds of km~s$^{-1}$) to propagate across 
the 126~pc separation to arrive, compress, and finally prompt the formation of stars within 
GAL\,110$-$13.  Additional evidence in support of this supernova scenario comes from the B5V star, 
HD\,201910, which is supposed to be a runaway star kicked out from a binary system in Lac\,OB1b, when   
one of the component stars became a supernova \citep{bla61,gie86}.   If this is so, the 
kinematic time scale of the star, 2.7~Myr, suggests that a supernova explosion occurred some 
2.7~Myr ago and the associated shocks subsequently caused GAL\,110$-$13 to develop its present 
cometary shape.

An alternative explanation is due to compression by ionization fronts from a massive star, which 
would be a less destructive method for star formation than a supernova explosion \citep{lef02}.  
We propose a scenario in which 10\,Lac---still in existence now---was born at the edge of the Lac 
molecular cloud, similar to that presented in Figure~\ref{fig:lac}, but with the cloud originally 
being more extended toward 10\,Lac.  Soon after its birth, 10\,Lac ionized the surrounding 
molecular clouds, exposing itself to the intercloud medium.  Assuming that most of the UV 
photons of 10\,Lac shortward of the Lyman limit were used to ionize the intercloud medium, then given 
a typical intercloud material density $\sim 0.2$~cm$^{-3}$ \citep{spi98,dys97}, the I-fronts 
would travel the 126~pc distance from 10\,Lac to GAL\,110$-$13 in about 2~Myr, a time scale still 
shorter than the main sequence life time of $\sim 3.6$~Myr of 10\,Lac \citep{sch97}.  
Regardless of which scenario actually happened, a supernova shock front or an ionization front, 
Lac\,OB1b is likely responsible for the creation of GAL\,110$-$13 and the associated stellar group.

\subsection{STAR FORMATION HISTORY IN LAC\,OB1}

In LBN\,437, star 52 (V375\,Lac) is the only young star located at the edge of the Lac molecular 
cloud, and interestingly there is no CTTS or HAeBe candidate behind the interaction region.  To 
check whether any PMS stars could have escaped the 2MASS detection limit of $J=15$~mag as a 
result of excessive dust extinction, we created the $E(B-V)$ map of the Lac molecular cloud 
\citep{sch98} (as done for the BRCs in Ori\,OB1; see Paper~I).  It is assumed that the $J$-band 
luminosities of the embedded PMS stars are the same as those of the visible PMS stars outside the 
cloud, namely the IRAS\,22343$+$3944 group.  Like other BRCs we have analyzed the overall 
extinction in the Lac molecular cloud is low, and the probability of nondetection is 0.014.  This 
means that there are indeed no embedded PMS stars and hence no ongoing star formation in the Lac 
molecular cloud.

The elongated Lac molecular cloud associated with LBN\,437 (Fig.~\ref{fig:lac}) may be just the 
remnant of a molecular cloud, which was originally perhaps larger, extending as far as to 10\,Lac.  
Upon the birth of 10\,Lac, its energetic photons evaporated and compressed the 
cloud, shaping the cloud into a pillar, similar to the case of GAL\,110$-$13.
The IRAS\,22343$+$3944 group and star 52 were then born on the compressed side of the cloud.  
At least 3 stars in the IRAS\,22343$+$3944 group exhibit 
forbidden lines, which is suggestive of their youth.  Star~52 is likely even younger because it is 
the exciting source of an HH outflow.  Apparently star 52 is the latest product 
in the star formation sequence by 10\,Lac in this cloud.   

\citet{ode92} derived a 30\% star formation efficiency for GAL\,110$-$13.  This is much higher 
than that of the few percent typical in star-forming regions \citep{whi95}.  Extinction is low in 
GAL\,110$-$13, with an A$_{J}$ less than 0.48~mag, as estimated from its $E(B-V)$ values, so the 
cloud is insufficiently dense to hide from our detection any embedded young stars similar to 
star 40.  As in the case for Ori\,OB1 (Paper~I), the BRCs in Lac\,OB1 also tend to have a 
relatively low dust extinction.  Such a low density condition is unfavorable for spontaneous,  
global cloud collapse.  Star formation however could take place at the interaction layer 
(the bright rim) of a molecular cloud.  A stellar group could form, such as witnessed in the 
IRAS\,22343+3944 and GAL\,110$-$13 groups. 

\begin{figure}
\epsscale{1.0}
\includegraphics[angle=90,scale=0.8]{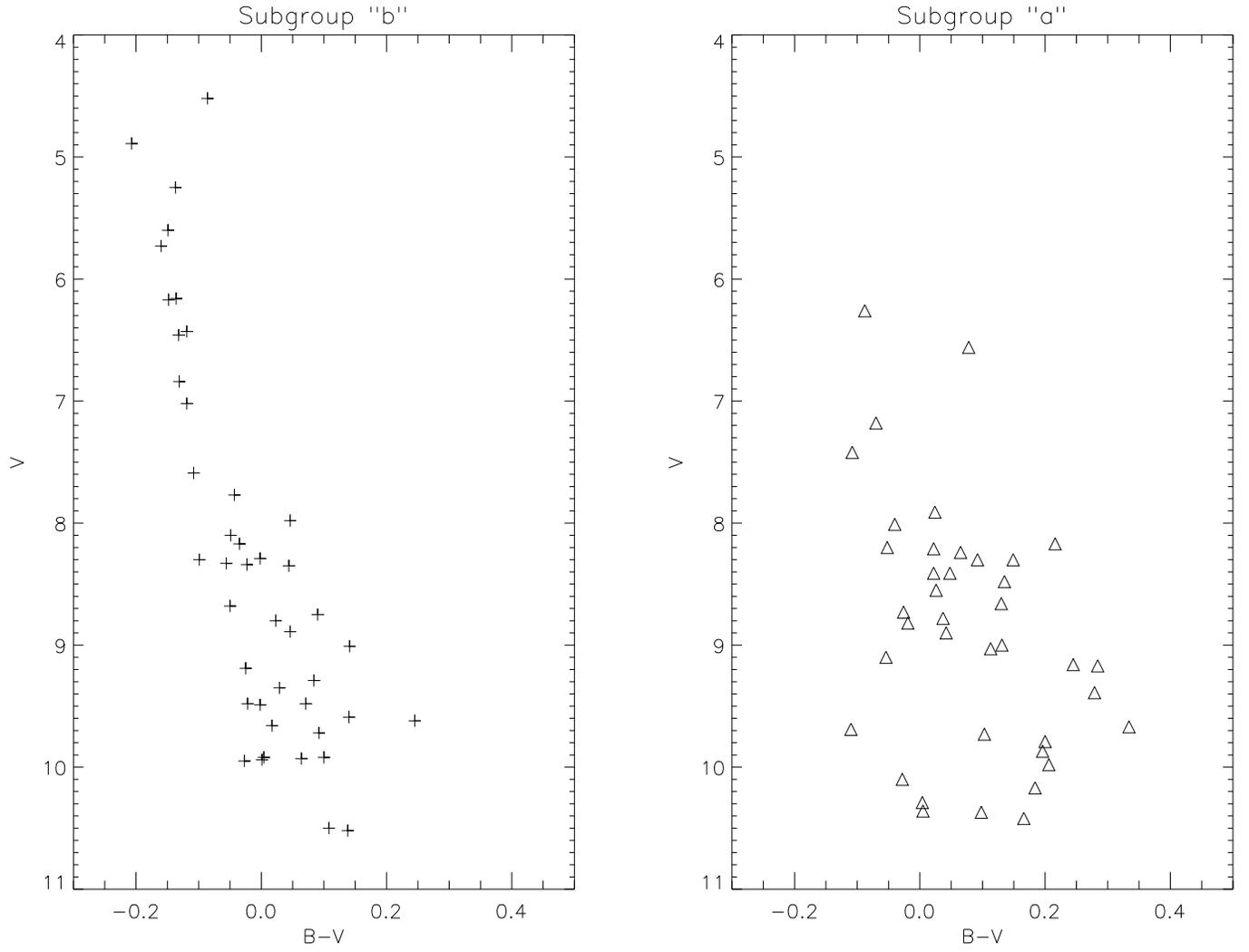}
\caption{Color-magnitude diagrams of the subgroups Lac\,OB1a and Lac\,OB1b.  The stars in 
Lac\,OB1b (pluses) form a clear main sequence, whereas those in Lac\,OB1a (triangles) are scattered 
to the right of the sequence, implying a younger age for Lac\,OB1a.}
\label{fig:cmd}
\end{figure}

\citet{bla58} and \citet{bla64,bla91} derived the ages of Lac\,OB1a and Lac\,OB1b, on the basis of 
stellar proper motions and radial velocities: 16--25~Myr and 12--16~Myr, respectively.  Both 
these ages are too old to be consistent with the existence of 10\,Lac (with a lifetime of less 
than $\sim 3.6$~Myr) and the CTTSs (typically aged a few Myr) in the region.  Thus Lac\,OB1a and 
Lac\,OB1b could not have formed at the same place and at the same time, because with a typical 
velocity dispersion of a few kilometers per second for an OB association \citep{dez99,deb99}, the 
two subgroups could not traverse the distance of 30--80~pc now between them.  We propose that both 
Lac\,OB1a and Lac\,OB1b are no more than a few Myr old, and Lac\,OB1a is younger than Lac\,OB1b.  
Figure~\ref{fig:cmd} shows the color-magnitude diagrams reconstructed from \citet{dez99} for the 
two subgroups.  It can be seen that the stars in Lac\,OB1b form a clear main sequence, whereas 
those in the subgroup Lac\,OB1a are widely scattered to the right of the sequence.  Some stars in 
Lac\,OB1a may well still be in the PMS phase, hence we postulate a younger age for Lac\,OB1a than 
for Lac\,OB1b.

It is possible that Lac\,OB1b was formed first after which the expanding I-fronts from Lac\,OB1b 
triggered new generations of stars along the Lac molecular cloud, the IRAS\,22343+3944 group 
and star 52.  A subsequent supernova or I-front then initiated the formation of stars in 
Lac\,OB1a; eventually the ``birth wave" reached GAL\,110$-$13.

\section{STAR FORMATION IN OB ASSOCIATIONS}

Triggered star formation has been suggested to have occurred close to \ion{H}{2} regions 
\citep{hes04,hes05}.  Our study finds clear chronological and positional causality of such 
processes on larger scales.  In $\lambda$ Ori, Ori\,OB1 and Lac\,OB1, we see supporting evidence 
of triggered star formation.  The UV photons from an O star create expanding I-fronts which 
evaporate and compress nearby molecular clouds, thereby shaping the clouds into BRCs or 
comet-shaped clouds.  The next generation of stars can then form efficiently, perhaps in 
groups, out of the compressed material.  The resulting newly formed stars would line up between 
the massive star and the molecular coulds in a formation and hence age sequence.  Stars at least 
as massive as late Herbig Be types could be formed via this process (see Table~4).  These stars would 
reach the main sequence with even earlier spectral types.  Triggered star formation could 
therefore produce not only low-mass stars, but also intermediate-mass or even massive stars.  In 
our sample, the HAeBe stars and CTTSs seem to be distributed spatially differently relative to a 
BRC, in the sense that the CTTSs tend to be located near the surface of a BRC, whereas the HAeBe 
stars appear preferentially to reside deeper into a BRC(e.g., star 44 in B\,30, star 52 in 
LBN\,437, and star 41 in LDN\,1616).

What we see in $\lambda$ Ori, Ori\,OB1, and Lac\,OB1 is in contrast to the scenario proposed by 
\citet{elm77} and \citet{lad87} for which massive stars are formed in shocked cloud layers by 
triggering, whereas low-mass stars are formed spontaneously via cloud collapse and fragmentation.  
A global cloud collapse would lead to starbirth spreading throughout the cloud, but this was 
not observed in our study.  Instead, no young stars are found embedded in clouds far behind the 
I-fronts.  More than mere ``fossil" molecular clouds, the BRCs present convenient snapshots of 
how star formation must have 
proceeded in an OB association.  When prompted to form, massive stars appear to favor denser 
environments where photoevaporation is relatively weak.  In comparison, when a dense 
molecular core near the ionization layer (i.e., current cloud surface) collapses, the accretion 
process has to compete with the mass loss arising from photoevaporation, leading to the formation 
of less massive stars or even substellar objects \citep{whi04}.  As the I-fronts progress, the 
remnant cloud is eventually dispersed, with stars of different masses remaining in the same 
volume.  Low- and intermediate-mass young stars in bright-rimmed or comet-shaped clouds on the 
border of an OB association are more likely to be formed by triggering.  Assuming a shock speed of  
$\sim 10$~km~s$^{-1}$, this would result in an age spread of several Myrs between member stars or 
star groups formed in the sequence.  If the velocities of the shocks are higher, as in the case of 
a supernova explosion, the age spreads would be less.


\section{CONCLUSIONS}

We first selected CTTSs and HAeBe stars in $\lambda$ Ori, Ori\,OB1, and Lac\,OB1 based on the 
2MASS colors.  These PMS stars are then utilized to trace recent star-forming activities.  The 
young stars are found to be lined up in an age sequence between massive stars and comet-shaped clouds or 
bright-rimmed clouds, with the youngest stars located near the cloud 
surfaces.  There are no PMS stars far behind the I-fronts.  These results support the scenario by 
which the Lyman continuum photons of a luminous O star create expanding I-fronts that would 
cause the evaporation and compression of nearby clouds to form BRCs or comet-shaped clouds, 
thereby inducing the birth of low- and intermediate-mass stars.  The BRCs provide us with a 
convenient setting in which to see that the HAeBe stars tend to form in the inner, denser parts of 
a cloud, whereas the CTTSs are formed near the photoevaporating cloud layers.  Young stars in 
bright-rimmed or comet-shaped clouds near a massive star are likely to have been 
formed by triggering.  Assuming a shock speed of $\sim 10$~km~s$^{-1}$, this would result in an age 
spread of several Myrs between member stars or star groups formed in the sequence.



\acknowledgments

We want to particularly thank Richard F. Green, Director of KPNO, who kindly provided us the 
director's discretionary time to accomplish this work.  We are also grateful to the staff of the 
Beijing Astronomical Observatory for their assistance during our observation runs and to the 
referee, Hans Zinnecker, for his suggestions helpful in improving the quality of this paper.  This 
research makes use of data products from the Two Micron All Sky Survey, which is a joint project 
of the University of Massachusetts and the Infrared Processing and Analysis Center/California 
Institute of Technology, funded by the National Aeronautics and Space Administration and the 
National Science Foundation (NSF).  We also used the Southern H-Alpha Sky Survey Atlas (SHASSA), 
supported by the NSF.  We acknowledge the financial support of grant NSC92-2112-M-008-047 from the 
National Science Council and 92-N-FA01-1-4-5 from the Ministry of Education of Taiwan.


\clearpage


\begin{deluxetable}{ccc}
\tablecaption{REGIONS STUDIED} 
\tablenum{1} 
\tablewidth{0pt}
\tablehead{
\colhead{Region} & \multicolumn{2}{c}{Approximate Coordinates} \\
} 
\startdata
\object{Lac\,OB1}$^{a}$      & $\ell \sim 83\arcdeg$ to $ 112\arcdeg$      
                                        & $b \sim -3\fdg 5$ to $-25\fdg 7 $                    \\
\object{Trapezium}$^{b}$     & RA  $ \sim 5^{h} 03^{m}$ to $ 5^{h} 32^{m}$ 
                                        & DEC $ \sim -1\arcdeg 45\arcmin$ to $ -8\arcdeg 10\arcmin$  \\
\object{$\lambda$ Ori}$^{c}$ & RA  $ \sim 5^{h} 23^{m}$ to $ 5^{h} 52^{m}$ 
                                        & DEC $ \sim +6\arcdeg 40\arcmin$ to $+14\arcdeg 22\arcmin$   \\
\object{Ori\,East}           & RA  $ \sim 5^{h} 52^{m}$ to $ 5^{h} 57^{m}$ 
                                        & DEC $ \sim +1\arcdeg 15\arcmin$ to $ +2\arcdeg 15\arcmin$   \\
Control Field 1                  & $\ell \sim 192\arcdeg$ to $ 260\arcdeg$     
                                        & $b  \sim +15\arcdeg$ to $ +44\arcdeg$                         \\
Control Field 2                  & RA $ \sim 20^{h} 24^{m}$ to $ 21^{h} 05^{m}$ 
                                        & DEC $ \sim +25\arcdeg 16\arcmin$ to $ +32\arcdeg 55\arcmin$   
\enddata
\tablenotetext{a}{Including BRC \object{LBN\,437} and comet-shaped cloud \object{GAL\,110$-$13} }
\tablenotetext{b}{Including BRCs, \object{IC\,2118}, \object{LDN\,1616} and \object{LDN\,1634} }
\tablenotetext{c}{Including BRCs, \object{B\,30} and \object{B\,35}  } 
\end{deluxetable}


\begin{deluxetable}{ccccc}
\tablecaption{Imaging Observations}
\tablenum{2}
\tablewidth{0pt}
\tablehead{\colhead{Fields}&\colhead{RA}&\colhead{DEC}&\colhead{Filter}&\colhead{Total Exp. Time } \\
                           &\colhead{ (J2000)} & \colhead{(J2000)} &   &  (s)                            
}
\startdata
B30         & 05:29:51.4 & +12:13:58 & H$\alpha$       & 5400 \\
B35         & 05:44:20.0 & +09:10:40 & H$\alpha$       & 5400 \\
Ori East    & 05:53:58.6 & +01:40:37 & H$\alpha$       & 3600 \\
LDN\,1616   & 05:07:06.0 & $-03:17:54$ & H$\alpha$       & 7200 \\
LDN\,1634   & 05:20:16.0 & $-05:49:28$ & H$\alpha$       & 3600 \\
IC\,2118    & 05:07:44.0 & $-06:12:35$ & H$\alpha$       & 2400 \\
LBN\,437    & 22:34:31.0 & +40:37:44 & H$\alpha$       & 3600 \\
LBN\,437    & 22:34:31.0 & +40:37:44 & [\ion{S}{2}] & 7200 \\
\enddata
\end{deluxetable}

\clearpage
\thispagestyle{empty}
\begin{deluxetable}{lccccl}
\tablecaption{CTTS and CTTS Candidates}
\tabletypesize{\footnotesize}
\rotate
\tablenum{3}
\tablewidth{0pt}
\tablehead{
\colhead{Star$^{a}$}&\colhead{2MASS}&\colhead{Emission Line(s)$^{b}$}&\colhead{Li$^{c}$}&\colhead{Obs.$^{d}$}&\colhead{Remarks}\\
}

\startdata
1  & J05065464-0320047 & H(-50.6), O(-0.6), Ca, He               & a & K    & \object{LkHa 333}, associated with LDN\,1616  \\ 
2  & J05073016-0610158 & H(-92.5), O(-10.8), S(-1.4), Fe, Ca, He & a & K    & associated with IC\,2118 \\ 
3  & J05073060-0610597 & H(-23.3), Ca?                           & a & K    & associated with IC\,2118 \\ 
4  & J05122053-0255523 & H(-12.1)                                & a & K    & \object{V531 Ori} \\ 
5  & J05141328-0256411 & H(-210.6), O(-2.5), Fe, Ca, He          & n & K    & \object{Kiso A-0975 16} \\ 
6  & J05152683-0632010 & H(-0.3)                                 & a & K    & H$\alpha$ emission is week,could be a WTTS \\ 
7  & J05162251-0756503 & H(-37.7), O(-1.0), Ca, He               & a & K    &  \\ 
8  & J05181685-0537300 & H(-57.3), O(-2.3), Fe, Ca, He           & n & K    & \object{Kiso A-0975 43} \\ 
9  & J05191356-0324126 & H(-52.1), Ca, He                        & a & K    & \object{Kiso A-0975 45} \\ 
10 & J05191549-0204529 & H(-10.7), O(-3.6), Ca                   & a & K    &  \\ 
11 & J05201945-0545553 & H(-26.9), Ca, He                        & a & K    & \object{Kiso A-0975 52}, IRAS\,05178-0548, 
                                                                     associated with LDN\,1634 \\ 
12 & J05202573-0547063 & H(-100.2), O?, Fe, Ca, He               & a & K    & \object{V534 Ori}, associated with LDN\,1634 \\ 
13 & J05203142-0548247 & H(-19.3), Ca, He                        & a & K    & \object{StHA 39}, associated with LDN\,1634 \\ 
14 & J05253979-0411020 & H(-138.9), Fe, Ca, He                   & n & K    & \object{Kiso A-0975 86} \\ 
15 & J05262158+1131339 & H(-15.4), O(-1.8), Fe, Ca               & - & B    & \object{IRAS 05235+1129} \\ 
16 & J05292393+1151576 & H(-40.5), Ca, He                        & a & B, K & \object{V649 Ori}, associated with B\,30 \\ 
17 & J05300203+1213357 & H(-33.6), Ca, He                        & a & B, K & \object{GX Ori}, IRAS\,05272+1211, associated with B\,30 \\ 
18 & J05301313+1208458 & H(-5.7), Ca                             & a & B, K & \object{GY Ori}, associated with B\,30 \\ 
19 & J05311615+1125312 & H(-21.9)                                & n & B, K & \object{V449 Ori} \\ 
20 & J05315128+1216208 & H(-127.3), O(-10.2), He                 & a & B, K & associated with B\,30 \\ 
21 & J05323207+1044178 & H(-96.7), Ca, He                        & - & B    &  \\ 
22 & J05324305+1221083 & H(-13.9), O(-1.3), Ca, He               & a & B, K & \object{V460 Ori}, IRAS\,05299+1219, associated with B\,30 \\ 
23 & J05330207+1137114 & H(-176.8), Fe, Ca, He                   & - & B    &  \\ 
24 & J05391268+0915522 & H(-215.6), Ca, He                       & - & B    &  \\ 
25 & J05432091+0906071 & H(-21.8), O(-0.5), Ca, He               & a & B, K & \object{V625 Ori}, IRAS\,05406+0904, associated with B\,35 \\ 
26 & J05440899+0909147 & H(-44.5), O(-2.1), Fe, Ca, He           & a & B, K & \object{QR Ori}, IRAS\,05413+0907, associated with B\,35 \\ 
27 & J05451493+0721223 & H(-5.6)                                 & n & B, K & \object{V661 Ori} \\ 
28 & J05452235+0904123 &                                         & - & B    & \object{FU Ori}, IRAS\,05426+0903, associated with B\,35 \\ 
29 & J05515035+0821066 & H(-181.9), Ca                           & a & B, K &  \\ 
30 & J05534090+0138140 & H(-29.0), O(-2.7), He                   & - & B    & \object{LkHA 334}, IRAS\,F05510+0137, associated with Ori East \\ 
31 & J05535869+0144094 & H(-37.0), Ca, He                        & - & B    & \object{LkHA 335}, IRAS\,F05513+0143, associated with Ori East \\ 
32 & J21370366+4321172 & H(-174.8), Fe, Ca, He                   & a & B, K & \object{V1082 Cyg} \\ 
33 & J21395545+4313082 & H(-83.6), Ca                            & - & B    &  \\ 
34 & J21535750+4659443 & H(-51.7)                                & - & B    & \object{LkHA 256} \\ 
35 & J22361978+4006273 & H(-63.9), O(-1.8), Ca                   & - & B    & associated with IRAS\,22343+3944 group \\ 
36 & J22362779+3954066 & H(-18.4)                                & - & B    & associated with IRAS\,22343+3944 group \\ 
37 & J22370328+4005185 & H(-10.3), Ca?, He?                      & a & B, K & associated with IRAS\,22343+3944 group \\ 
38 & J22371683+3952260 & H(-130.8), O(-4.5), Ca                  & - & B    & associated with IRAS\,22343+3944 group \\ 
39 & J23104483+4508511 & H(-7.9)                                 & n & B, K &  \\ 
40 & J23373847+4824119 & H(-19.2)                                & a & B, K & \object{BM And}, associated with GAL\,110$-$13 \\ 

\enddata
\tablenotetext{a}{Star 1--31 and 32--40 are in the Orion and Lacerta regions, respectively.}
\tablenotetext{b}{H--H$\alpha$, Ca--\ion{Ca}{2} (K, H (3934, 3968~\AA), and/or 
infrared triplet (8498, 8542, 8662~\AA)), He--\ion{He}{1} (5876~\AA), 
O--[\ion{O}{1}] (6300~\AA), S--[\ion{S}{2}] (6717~\AA), Fe--\ion{Fe}{2} 
(4924~\AA), The number following H, O, S, are the equivalent widths of H$\alpha$, [\ion{O}{1}], 
and [\ion{S}{2}], respectively.}
\tablenotetext{c}{a--absorbtion, n--no absorbtion, - --low spectral resolution in BAO}
\tablenotetext{d}{B--BAO, K--KPNO}
\end{deluxetable}

\clearpage


\begin{deluxetable}{lccccl}
\tablecaption{Herbig Ae/Be Stars}
\tabletypesize{\footnotesize}
\rotate
\tablenum{4}
\tablewidth{0pt}
\tablehead{
\colhead{Star$^{a}$}&\colhead{2MASS}&\colhead{Emission Line(s)}&\colhead{Sp. Type$^{b}$}&\colhead{Obs.$^{c}$}&\colhead{Remarks}\\
}
\startdata

41 & J05042998-0347142 & H       & A3e & K    & \object{UX Ori}, IRAS 05020-0351, associated with LDN\,1616 \\
42 & J05113654-0222484 & H       & A3e & K    &  \\
43 & J05305472+1421524 & H       & F2e & K    &  \\
44 & J05312805+1209102 & H, O    & A2e & K    & \object{HK Ori}, IRAS\,05286+1207, associated with B\,30 \\
45 & J05313515+0951553 & H       & B9e & K    & \object{IRAS 05288+0949} \\
46 & J05315724+1117414 & H       & A0e & B    & \object{HD 244604}, IRAS\,05291+1115 \\
47 & J05350960+1001515 & H, O?   & B9e & B    & \object{V1271 Ori}, IRAS\,05324+0959 \\
48 & J05390921+0925301 & H       & F7e & B, K & \object{V506 Ori} \\
49 & J21462666+4744154 & H, O    & B9e & K    &  \\
50 & J21514726+4615115 & H       & A9e & K    & \object{LR Cyg} \\
51 & J22154039+5215559 & H       & A2e & B    &  \\
52 & J22344101+4040045 & H, O, S & A2e & K    & \object{V375 Lac} \\
53 & J22363511+4000156 & H, O    & B8e & B    & associated with IRAS\,22343+3944 group \\

\enddata
\tablenotetext{a}{Star 41--48 and 49--53 are in the Orion and Lacerta regions, respectively.}
\tablenotetext{b}{H--H$\alpha$, O--[\ion{O}{1}] (6300~\AA), S--[\ion{S}{2}] 
(6717~\AA)}
\tablenotetext{c}{B--BAO, K--KPNO}
\end{deluxetable}

\clearpage


\begin{deluxetable}{lcccl}
\tablecaption{Non-PMS Stars}
\tabletypesize{\footnotesize}
\rotate
\tablenum{5}
\tablewidth{0pt}
\tablehead{
\colhead{Star}&\colhead{2MASS}&\colhead{Sp. Type}&\colhead{Obs.$^{a}$}&\colhead{Remarks}\\
}
\startdata

54 & J05232026+0934432 & A0 & B, K & \object{TYC 704-1857-1} \\ 
55 & J05285405-0606063 & Me & K    & \object{Kiso A-0975 119}, IRAS\,05264-0608 \\ 
56 & J05413010+1418225 & C  & K    & \object{BC 203} \\ 
57 & J05442880+0652019 & M  & B    &  \\ 
58 & J05464207+0643469 & C  & B    & \object{IRAS 05440+0642} \\ 
59 & J05480851+0954012 & Ce & B, K & \object{V638\,Ori}, IRAS\,05453+0953 \\ 
60 & J07323273+2647156 & C  & K    & object{FBS 0729+269} \\ 
61 & J07475919+2052254 & Ce & K    &  \\ 
62 & J08231037-0153257 & C  & K    &  \\ 
63 & J08292902+1046241 & C  & K    & \object{FBS 0826+109} \\ 
64 & J08423302+0621195 & M  & K    &  \\ 
65 & J08541870-1200541 & Ce & K    & \object{IRAS 08519-1149} \\ 
66 & J09111450-0922053 & Me & K    & \object{VV Hya} \\ 
67 & J09333061-2216282 & M  & K    &  \\ 
68 & J20245404+2609115 & M  & B    &  \\ 
69 & J20291739+2617284 & Me & B    & \object{IRAS 20271+2607} \\ 
70 & J20304177+2812340 & M  & B    & \object{DU Vul}, IRAS\,20285+2802 \\ 
71 & J20311267+2612270 & M  & B    &  \\ 
72 & J20415136+2752525 & M  & B    & \object{IRAS 20397+2742} \\ 
73 & J20532040+2516196 & C  & B    &  \\ 
74 & J20551307+3254065 & M  & B    &  \\ 
75 & J20555284+2640515 & M  & B    & \object{UY Vul}, IRAS\,20537+2629 \\ 
76 & J21040556+2632111 & M  & B    & \object{V444 Vul}, IRAS\,21019+2620 \\ 
77 & J21244172+4437134 & Ce & B    & \object{V1563 Cyg}, IRAS\,21228+4424 \\ 
78 & J21383182+4542469 & Ce & K    & \object{V1568 Cyg}, IRAS\,21366+4529 \\ 
79 & J21595030+3313596 & M  & B    &  \\ 
80 & J22024329+4216400 &BL Lac&K   & \object{BL Lac} \\ 
81 & J22055958+3530057 & M  & B    & \object{XX Peg} \\ 
82 & J22070988+2828374 & M  & B    & \object{V392 Peg}, IRAS\,F22048+2813 \\ 
83 & J22075421+4105113 & M  & B    & \object{V379 Lac}, IRAS\,22057+4050 \\ 
84 & J22084406+4855248 & M  & B    & \object{V426 Lac}, IRAS\,22067+4840 \\ 
85 & J22121336+4646065 & C  & B    & IRAS\,22101+4631 \\ 
86 & J22135091+2447203 & M  & B    &  \\ 
87 & J22213857+3335586 & C  & B    &  \\ 
88 & J22261658+4221089 & A0 & B    &  \\ 
89 & J22295650+4546539 & Ce & B    & \object{V386 Lac} \\ 
90 & J22313443+4816005 & C  & K    & \object{V387 Lac}, IRAS\,22294+4800 \\ 
91 & J22314368+4748038 & PN & K    & \object{PN G100.0-08.7}, IRAS\,22296+4732 \\ 
92 & J22451504+5051534 & Ce & B    & \object{HL Lac}, IRAS\,22431+5036 \\ 
93 & J22491976+5154487 & M  & B    & IRAS\,22472+5138 \\ 
94 & J22514566+4921137 & C  & B    & IRAS\,22495+4905 \\ 
95 & J22521809+3413364 & M  & B    & IRAS\,22499+3357 \\ 
96 & J22592372+4811589 & Me & K    &   \\ 
97 & J23023314+4649483 & M  & B    & \object{NSV 14395}, IRAS\,23002+4633  \\ 
98 & J23113005+4702525 & M  & B    & IRAS\,23092+4646 \\ 
99 & J23175960+4645122 & M  & B    & \object{AO And}, IRAS\,23156+4628 \\ 
\enddata
\tablenotetext{a}{B--BAO, K--KPNO}
\end{deluxetable}

\clearpage


\begin{deluxetable}{lll}
\tablecaption{Differences between triggered and spontaneous star formation}
\rotate
\tablenum{6}
\tablewidth{0pt}
\tablehead{\colhead{}&\colhead{Triggered}&\colhead{Spontaneous}}

\startdata
Sequential star formation  & Yes.  PMS stars close to triggering           & No                                             \\
                           & sources are older than those close to BRCs    &                                                \\
Stellar distribution       & PMS stars are located between triggering      & PMS stars can be anywhere, including being far \\
                           & sources or around surfaces of BRCs with no    & behind the surface of a BRC                    \\
                           & young stars embedded much behind the I-fronts &                                                \\
Star formation efficiency  & High                                          & Low, less than a few percent.                  \\
\enddata
\end{deluxetable}

\clearpage


\begin{deluxetable}{lcccccc}
\tablecaption{Proper Motions of Stars in GAL\,110$-$13 and Lac\,OB1 Subgroups a and b}
\tablenum{7}
\tablewidth{0pt}
\tablehead{
\colhead{Star}&\colhead{Sp. type}&\colhead{pmRA}    &\colhead{pmDEC}   &\colhead{e$\_$pmRA}&\colhead{e$\_$pmDEC} &\colhead{Reference}\\ 
\colhead{}    &\colhead{}        &\colhead{(mas/yr)}&\colhead{(mas/yr)}&\colhead{(mas/yr)} &\colhead{(mas/yr)} &\colhead{}
           } 
\startdata
HD\,222142 & B9.5 V    &  0.3 & -3.1 & 0.6 & 0.6 & UCAC2\\
HD\,222086 & B9 V      &  0.5 & -2.8 & 1.0 & 1.1 & UCAC2\\
HD\,222046 & B8 Vp     &  0.4 & -2.7 & 1.0 & 1.0 & UCAC2\\
Star 40    & Continuum &  3.4 & -7.8 & 2.7 & 2.6 & UCAC2\\
Lac\,OB1a  & -         & -0.3 & -3.7 & -   & -   & Hipparcos\\
Lac\,OB1b  & -         & -0.5 & -4.6 & -   & -   & Hipparcos\\
\enddata

\end{deluxetable}

\clearpage

\clearpage

\clearpage

\clearpage

\clearpage

\clearpage

\clearpage

\clearpage

\clearpage

\clearpage

\end{document}